\documentclass[aps,prd,preprint,showpacs,superscriptaddress,nofootinbib]{revtex4-1}
\usepackage[T1]{fontenc} 
\usepackage{pstricks}
\usepackage{color}
\usepackage{slashed}
\usepackage{graphicx,graphics}
\usepackage{ulem}
\usepackage{verbatim}
\usepackage{amssymb}
\usepackage{amsthm}
\usepackage{array}
\usepackage{amsmath,amsfonts}
\usepackage{caption}
\usepackage{subcaption}
\newcommand{\bea}{\begin{eqnarray}}
\newcommand{\eea}{\end{eqnarray}}

\newcommand{\be}{\begin{equation}}
\newcommand{\ee}{\end{equation}}

\newcommand{\eg}{{\it e.g.}}
\def\bsp#1\esp{\begin{split}#1\end{split}}
\def\bpm{\begin{pmatrix}}
\def\epm{\end{pmatrix}}

\def\W{{\tilde W}}

\def\B{{\tilde B}}

\def\d{\mathrm{d}}
\def\tchi{{\tilde \chi}}
\begin{document}

\preprint{{\tiny CUMQ/HEP 184, HIP-2014-18/TH, HRI-RECAPP-2014-019} }
\title{\Large  Left-right supersymmetry after the Higgs boson discovery}

\author{Mariana Frank}
\email{mariana.frank@concordia.ca}
\affiliation{Department of Physics, Concordia University, 7141 Sherbrooke St.
  West, Montreal, Quebec, Canada H4B 1R6}
\author{Dilip Kumar Ghosh}
\email{tpdkg@iacs.res.in}
\affiliation{Department of Theoretical Physics, 
Indian Association for the   Cultivation of Science, 2A \& 2B Raja S.C. Mullick Road,
Kolkata 700 032,
India}
\author{Katri Huitu}
\email{katri.huitu@helsinki.fi}
\affiliation{Department of Physics and Helsinki Institute of Physics, P.O. Box 64 (Gustaf H\"allstr\"omin katu 2), FIN-00014 University of Helsinki, Finland}
\author{Santosh Kumar Rai}
\email{skrai@hri.res.in}
\affiliation{Regional Centre for Accelerator-based Particle Physics, 
Harish-Chandra Research Institute,  Chhatnag Road, Jhusi, Allahabad 211019, 
India}
\author{Ipsita Saha}
\email{tpis@iacs.res.in}
\affiliation{Department of Theoretical Physics, 
Indian Association for the   Cultivation of Science, 2A \& 2B Raja S.C. Mullick Road,
Kolkata 700 032, India}
\author{Harri Waltari}
\email{harri.waltari@helsinki.fi}
\affiliation{Department of Physics and Helsinki Institute of Physics, P.O. Box 64 (Gustaf H\"allstr\"omin katu 2), FIN-00014 University of Helsinki, Finland}

\begin{abstract}
We perform a thorough analysis of the parameter space of the minimal left-right supersymmetric model in agreement with the LHC data. The model contains left- and right-handed fermionic doublets, two Higgs bidoublets, two Higgs triplet representations, and one singlet, insuring a charge-conserving vacuum. We impose the condition that the model complies with the experimental constraints on supersymmetric particles masses and on the doubly-charged Higgs bosons, and require that the parameter space of the model satisfy the LHC data on neutral Higgs signal strengths at $2\sigma$. We choose benchmark scenarios by fixing some basic parameters and scanning over the rest. The LSP in our scenarios is always the lightest neutralino. We find that the signals for $H\to \gamma \gamma$ and $H \to VV^\star$ are correlated, while $H \to b \bar b$ is anti-correlated with all the other decay modes, and also that the contribution from singly-charged scalars dominate that of the doubly-charged scalars in $H\to \gamma \gamma$ and $H \to Z\gamma$ loops, contrary to Type-II seesaw models. We also illustrate the range for mass spectrum of the LRSUSY model in light of planned measurements of the branching ratio of $H\to \gamma \gamma$ to 10\% level.
\end{abstract}

%\keywords{LHC phenomenology, Left Right Symmetry, Supersymmetry}

\maketitle

%\flushbottom

%%%%%%%%%%%%%% Begin Main Part %%%%%%%%%%%%%%%%%%%%%%%%%%%%%%%%%%%%%%%%%
 \section{Introduction}
 \label{sec:intro}
 %%%%%%%%%%%%%%%%%%%
The discovery of the Higgs boson at the LHC \cite{Chatrchyan:2012ufa, Aad:2012tfa} highlighted the importance of the search for signs for 
Beyond the Standard Model (BSM) in the Higgs boson decay or production modes.
While the amount of data at LHC with $\sqrt{s}=7$ and $8$ TeV is still rather limited, if combined with the measured Higgs mass, it seems to fit very well the Standard Model (SM) predictions and thus 
it allows one to restrict the parameter space of many BSM models. 
 In the SM, the Higgs signal rates (cross section times branching ratios) are completely fixed by the Higgs mass. Precise predictions for decays into various channels can be combined with the experimental data to define a signal strength for each decay $\mu_i$, normalized such that the SM corresponds to $\mu_i=1$. The mass of the Higgs boson is measured using the decay modes with clear mass peaks,  
$H \rightarrow ZZ^* \rightarrow  4\ell$ and $H \rightarrow \gamma\gamma$ 
\cite{ATLAS:2012ad,Chatrchyan:2012tw}.
Other decay modes of Higgs into $WW^*$, $b\bar b$, and $\tau^+ \tau^-$ has been measured both by CMS and ATLAS \cite{Chatrchyan:2012ufa, Aad:2012tfa}.
For recent updates in these channels we refer to \cite{ATLAS:2014,CMS:2014,Khachatryan:2014ira}.

There are two ways
in which a non-minimal BSM Higgs sector could be revealed experimentally: either
directly through the discovery of additional scalar states,  or through precise measurements of
the Higgs properties, that would  indicate deviations from the SM predictions
for the scalar state discovered at 126 GeV. At present, the only  measurement in the Higgs sector, which, within the accuracy of measurement, seems to possibly differ
from the SM prediction, is the di-photon mode as measured by the ATLAS \cite{ATLAS:2014} and CMS \cite{CMS:2014,Khachatryan:2014ira} Collaborations. 
Consequently, there has been  a lot of recent work on the decay rate of the Higgs boson to two photons. 

One approach would be to fit the data in a model-independent way \cite{Flacher:2008zq,Baak:2013ppa,Giardino:2013bma,Belanger:2013xza}.  One could also evaluate Higgs decays in  specific BSM scenarios, which lead to the  modifications of the Higgs couplings, especially for cases where additional particles which interact with the Higgs boson  exist, and/or where there is an extended Higgs sector. In the Standard Model the dominant contribution to the decay $H \rightarrow\gamma\gamma$ is given by the interference between the top quarks and the W-boson contributions at one-loop level.
In a number of possible frameworks for BSM physics, many more particles are present in the loop, and their contribution  can interfere either constructively or destructively with the loops containing the SM particles. The quest for the identity of the particle discovered at the LHC has thus taken the approach of focusing on a specific new Higgs model, including models with extended gauge and/or Higgs sector and exotic particles,  without supersymmetry \cite{ Arhrib:2012ia,Swiezewska:2012eh, Draper:2012xt, Arbabifar:2012bd, Akeroyd:2012ms, Picek:2012ei,Almeida:2012bq,Chiang:2012qz,Carena:2012xa},  in MSSM or extended supersymmetric models \cite{Bandyopadhyay:2014tha,Delgado:2012sm,  Kitahara:2012pb,Basso:2012tr}, or in extra-dimensional models \cite{Flacke:2013nta}. 
In addition, the effects of BSM physics can appear in the changes of partial widths of other known Higgs decay modes. Already,  a small change in the dominant decay mode to $b$-quarks can significantly change the di-photon rate. The BSM models can also generate new decay modes of Higgs bosons, {\it e.g.} an invisible decay mode,  possible in some models and experimentally allowed.

On fundamental theoretical grounds, it is expected that the Standard Model is not the final theory, and supersymmetric models remain good candidates for  realistic BSM models.
The minimal supersymmetric standard model (MSSM), favored because it is the simplest supersymmetric extension of the SM, is by far the most studied version of the
supersymmetric models. 
However, in view of the LHC measurements, it is itself somewhat fine tuned, and it inherits some of the  
 limitations of the SM as well, such as the absence of neutrino masses.
In particular, one imposes by hand a discrete symmetry, $R$-parity, to prevent the fast decay 
of proton without invoking very small Yukawa-type couplings.
This can be cured by including in the gauge group a part which automatically takes care of the
conservation of $R$-parity, $R_p=(-1)^{B-L+2s}$, where $B$ is the baryon number, $L$ is the lepton number,
and $s$ is spin.
Every model which contains $U(1)_{B-L}$ as part of the gauge symmetry, conserves $R$-parity
at the Lagrangian level, and thus it can be only spontaneously broken, if at all.

Left-right supersymmetric (LRSUSY) models are based on enlarging the SM symmetry to the
$SU(3)_c \otimes SU(2)_L \otimes SU(2)_R \otimes U(1)_{B-L}$ gauge group  \cite{Francis:1990pi,Huitu:1993uv, Huitu:1993gf}. 
The models  emerge from breaking of gauge unification scenarios such as $SO(10)$ or $E_6$ \cite{Babu:1993we, Babu:1994dq, Frank:1999wz, Frank:1999ys, Mohapatra:1997sp}, or as custodial symmetry in extra-dimensional models, in particular the warped space models \cite{Agashe:2003zs,Agashe:2006at}.
The LRSUSY models can contain triplet scalars, which by interacting with leptons generate
masses for neutrinos via seesaw mechanism \cite{Mohapatra:1979ia}.
The measured oscillations \cite{Cleveland:1998nv, Fukuda:1998mi} between neutrino
generations would not be possible without nonzero mass of the neutrinos, which can be taken as an 
experimental evidence for the BSM physics.
The LRSUSY model also resolves some other problems plaguing the MSSM, {\it e.g.} a solution to 
the strong and electroweak (EW) $CP$ problems \cite{Mohapatra:1995xd,Mohapatra:1996vg, Kuchimanchi:1995rp, Kuchimanchi:1993jg}.
The triplet Higgs representations in the LRSUSY model contain both singly and doubly charged scalars and higgsinos, which would be expected to
modify the di-photon rate compared to the Standard Model.

It is our aim in this paper to study the Higgs sector, in particular the relationship between the mass parameters, supersymmetric spectrum and Higgs decay widths of the model. Our aim is two-fold: to show that the model can allow an enhancement of the Higgs branching ratio into two photons, and at the same time agree with the limits on other decay modes. Our second goal is,  based on the analysis of the Higgs sector, to restrict and/or make some general predictions about the parameter space of LRSUSY models. Our work is organized as follows. 
After reviewing in Sec.~\ref{sec:model} the parts of the model relevant for our purposes, we study the spectrum
of the model in Sec.~\ref{sec:cspectrum}.
In Sec.~\ref{sec:gamgam} we study the rare decay modes $H \rightarrow \gamma\gamma, \, Z\gamma$ to see if the
corresponding branching ratios differ from the SM.
In Sec.~\ref{sec:restrictions} we consider the restriction to the parameter space, and finally we conclude in
Sec.~\ref{sec:conclusion}.

%%%%%%%%%%%%%%%%%%%%%%%%%%%%%%%%%%%%
%%%% Model description
%%%%%%%%%%%%%%%%%%%%%%%%%%%%%%%%%%%% 
\section{The Higgs sector of the  left-right supersymmetric model} 
\label{sec:model}
%%%%%%%%%%%%%%%%%%%%%%%%%%%%%%%%%%%%
Left-right supersymmetric models are based on enlarging the SM symmetry to the
$SU(3)_c \otimes SU(2)_L \otimes SU(2)_R \otimes U(1)_{B-L}$ gauge group  \cite{Francis:1990pi,Huitu:1993uv, Huitu:1993gf}. 

The chiral (matter) 
sector of the theory contains left-handed ($Q_L$ and $L_L$) and right-handed
($Q_R$ and $L_R$) doublets of quark and lepton supermultiplets,
\be\bsp
    (Q_L)^{  i  } = \bpm u_L^{i }\\d_L^{i } \epm =
      \big({\bf 3}, {\bf 2},{\bf 1},\frac13\big) \ , 
 &\qquad 
    (Q_R)_{i} = \bpm u^{\mathbf c}_{R i } & d^{\mathbf c}_{R i } \epm=
      \big({\bf \bar 3}, {\bf 1},{\bf 2}^*,-\frac13\big) \ ,\\
    (L_L)^{ i} = \bpm \nu_L^i \\ \ell_L^i\epm =
     \big({\bf 1}, {\bf 2},{\bf 1},-1\big)\ ,
 & \qquad 
    (L_R)_{i} = \bpm \nu_{R i}^{\mathbf c} & \ell_{R i}^{\mathbf c} \epm =
     \big({\bf 1}, {\bf 1},{\bf 2}^*,1\big)\ ,
\esp \label{eq:fieldcontent} \ee
where the 
 $i$ is a generation index,   ${\mathbf c}$ denotes charge conjugation, and, for simplicity,  we have suppressed color indices.

The gauge sector of the theory includes  gauge and gaugino fields, corresponding to the four gauge groups:
\be \bsp
  SU(3)_c: ~V_c   =&\ ({\bf 8},{\bf 1},{\bf 1},0) \equiv 
   \big(g_\mu^a, \tilde g^a \big)\ , \\
  SU(2)_L :~ V_{2L}=&\ ({\bf 1},{\bf 3},{\bf 1},0) \equiv 
   \big(W_{L\mu}^k, \tilde W_L^k \big)\ , \\
  SU(2)_R:~ V_{2R}=&\ ({\bf 1},{\bf 1},{\bf 3},0) \equiv 
   \big(W_{R\mu}^k, \tilde W_R^k \big)\ , \\
  U(1)_{B-L}:~ V_{B-L}=&\ ({\bf 1},{\bf 1},{\bf 1},0) \equiv 
   \big( B_\mu, \tilde{ B} \big) \ .
\esp \label{eq:vectormultiplets}\ee
The 
 $SU(3)_c \otimes SU(2)_L \otimes SU(2)_R \otimes U(1)_{B-L}$ gauge group is
broken down to the Standard Model gauge group via a set of two $SU(2)_R$ Higgs
triplets $\Delta^c$ and $\bar \Delta^c$, which are evenly charged under the $B-L$ gauge
symmetry\footnote{Doublets are also possible, but triplets are preferred as they facilitate the seesaw mechanism for neutrino mass generation \cite{Mohapatra:1979ia}.}. Often, in non-minimal models, extra $SU(2)_L$ Higgs triplets $\Delta$ and
$\bar \Delta$ are introduced to preserve parity at  higher scales.
Unfortunately, with the triplet representation, the minimum of the scalar potential is charge-violating, unless  the
right-chiral scalar neutrinos get vacuum expectation values (vevs), breaking
$R$-parity spontaneously \cite{Kuchimanchi:1993jg,Huitu:1994zm}.  

Three 
scenarios have been proposed which remedy this situation.  First, to avoid $R$-parity violation,  in Refs.\ \cite{Aulakh:1998nn,Chacko:1997cm} non-renormalizable operators are introduced at Planck scale, which shift the minimum of the potential. Second,  in Refs.\
\cite{Babu:2008ep,Frank:2011jia}, an additional singlet chiral supermultiplet ($S$) is
added  to the field content of the model, leading, after including one-loop Coleman-Weinberg corrections, to an $R$-parity conserving minimum of
the scalar potential. And third, in Refs. \cite{Aulakh:1997fq, Aulakh:1997ba}, two
extra Higgs triplets  $\Sigma_1 =({\bf 1},{\bf 3},{\bf 1},0)$ and $\Sigma_2= ({\bf 1},{\bf 1},{\bf
3},0)$ are included, yielding symmetry breaking with
conserved $R$-parity at tree-level. In this work we adopt the second approach as the 
minimal solution. The breaking of the $SU(2)_L\otimes U(1)_Y$ symmetry to $U(1)_{\rm EM}$ is
achieved with two $SU(2)_L \otimes SU(2)_R$ Higgs
bidoublets $\Phi_1$ and $\Phi_2$ which also  generate
non-trivial quark mixing angles \cite{Babu:1998tm}. The field content
of the Higgs sector is thus summarized as
\be\bsp
  \Phi_1 = \bpm \phi_1^0&\phi_2^+\\ \phi_1^-&\phi^{0}_2 \epm =
     \big( {\bf 1}, {\bf 2}, {\bf 2}^*,0\big) \ ,
  & \qquad 
    \Phi_2 = \bpm \chi^{0}_1&\chi_2^+\\ \chi_1^-&\chi_2^0 \epm  =
     \big( {\bf 1},  {\bf 2}, {\bf 2}^*,0\big)\ , \\
     \Delta^c = 
      \bpm
        \frac{\delta^{c\,-}}{\sqrt{2}}  & \delta^{c\,0}\\
        \delta^{c\,--} & \frac{-\delta^{c\,-}}{\sqrt{2}} 
      \epm  = \big( {\bf 1},  {\bf 1}, {\bf 3},-2\big)\ , 
   & \qquad  \bar{\Delta}^c  = 
      \bpm
        \frac{\bar{\delta}^{c\,+}}{\sqrt{2}} & \bar{\delta}^{c\,++}\\
        \bar{\delta}^{c\,0} & \frac{-\bar{\delta}^{c\,+}}{\sqrt{2}} 
      \epm = \big( {\bf 1},  {\bf 1}, {\bf 3} ,2\big) \ , \\
       \qquad
      &\hspace{-1.2cm} S = \big({\bf 1}, {\bf 1},{\bf 1},0\big) \ .
\esp\label{eq:fieldcontent1}
\ee
The superpotential describing the interactions among the chiral
supermultiplets of the model is
\bea
W(\phi) &=&\nonumber
   (Q_L)^{i} Y_Q^1 (\Phi_1) (Q_R)_{i} +
   (Q_L)^{ i} Y_Q^2 ( \Phi_2) (Q_R)_{i} +
   (L_L)^i Y_L^1 ( \Phi_1) (L_R)_i + (L_L)^i Y_L^2 ( \Phi_2) (L_R)_i \\
&&+
  i (L_R)_i f (\Delta^c) (L_R)^i 
   + S \,\left[\lambda{\rm Tr}(\Delta^c \cdot  \bar\Delta^c)-{\cal M}_{R}^{2} \right]
   +  \lambda_{12} S \,{\rm Tr} (\Phi_1 \cdot {\Phi}_2)
   \label{eq:Wtrip}
   \eea
where $i$ is a generation index, $Y^j_Q$,  $Y^j_L$ ($j=1,2$), and $f$ are $3\times 3$ matrix Yukawa couplings.

The full scalar potential of the model, which is minimized to obtain the masses and composition of the Higgs bosons, is given by 
\begin{eqnarray}
V_{F}&=&\left|\lambda{\rm Tr}(
\Delta^c\bar\Delta^c)+\lambda_{12}{\rm Tr}(\Phi_{1}^{T}\Phi_{2}
)-{\cal M}_{R}^{2}\right|^2+ \lambda^{2}|S|^{2}\left|{\rm
Tr}(\Delta^c\Delta^{c\dagger})+{\rm
Tr}(\bar\Delta^c\bar\Delta^{c\dagger})\right|,
\nonumber\\
V_{\rm soft} &=& M_{1}^{2}{\rm Tr}
(\Delta^{c\dagger}\Delta^c)+
M_{2}^{2}{\rm
Tr}(\bar\Delta^{c\dagger}\bar\Delta^c)
+M_{3}^{2}\Phi_{1}^{\dagger}\Phi_{1}+M_{4}^{2}\Phi_{2}^{\dagger}
\Phi_{2}+M_{S}^{2}|S|^{2}\nonumber\\
&+& \{A_{\lambda}\lambda
S{\rm Tr}(\Delta^c\bar\Delta^c)-
C_{\lambda}{\cal M}_R^2 S + h.c.\}, \nonumber\\
V_D &=&\frac{g_{L}^{2}}{8}\sum_{i}\left|{\rm Tr}(\Phi_{1}
\Phi_{2}^{\dagger})\right|^2+
\frac{g_{R}^{2}}{8}\sum_{i}\left|{\rm Tr}(2 \Delta^{c\dagger}
\Delta^c+2\bar\Delta^{c\dagger} \bar\Delta^c+\Phi_{1}
\Phi_{2}^{\dagger})\right|^2\nonumber\\
&+& {\frac{g_{B-L}^{2}}{2}} \left|{\rm
Tr}(-\Delta^{c\dagger}\Delta^c+\bar\Delta^{c\dagger}\bar\Delta^c)\right|^2.
\end{eqnarray}
The gauge symmetry is spontaneously broken in two steps. First the $SU(2)_R \otimes
U(1)_{B-L}$ gauge group is broken to the SM gauge group, which is
subsequently broken to the electromagnetic group $U(1)_{\rm EM}$ by the vacuum expectation values (VEVs) of  the
neutral components of the Higgs fields
\be\bsp
   & \langle S \rangle = \frac{v_s}{\sqrt{2}} e^{i
    \alpha_s}\ ,
  \quad
  \langle \Phi_1 \rangle =  \bpm 
     \quad \frac{v_1}{\sqrt{2}} \quad & 0 \\ 
      0 & \frac{v_1^\prime}{\sqrt{2}}  e^{i \alpha_1} \epm \ , 
  \quad 
  \langle \Phi_2 \rangle = \bpm 
     \quad \frac{v_2^\prime }{\sqrt{2}} e^{i \alpha_2} \quad & 0\\ 
     0 &\frac{v_2}{\sqrt{2}} \epm \ , \\
  &
    \langle \Delta^c \rangle =  \bpm 0 & \frac{v_{R}}{\sqrt{2}} \\ 0 & 0 \epm , \
  \quad
  \langle \bar\Delta^c \rangle =  \bpm 0 & 0 \\ \frac{\bar{v}_{R}}{\sqrt{2}} & 0 \epm .
\esp\ee
The VEVs 
$v_{R}$, $\bar{v}_{R}$, $v_1$, $v_2$, $v^\prime_1$, $v^\prime_2$ and $v_s$ can be chosen real and
non-negative, while the only complex phases which cannot be rotated away by
means of suitable gauge transformations and field redefinitions are
denoted by explicit angles $\alpha_1$, $\alpha_2$ and $\alpha_s$. However, as the $CP$-violating $W^\pm_L-W^\pm_R$ mixing is proportional 
to $v_1 v_1^\prime e^{i \alpha_1}$ and $v_2 v_2^\prime e^{i
\alpha_2}$, and is constrained to be small by $K^0-{\bar K}^0$ mixing data,  this forces the angles to be very small.  As the number of degrees of freedom remains large, 
we assume the hierarchy
\be \label{eq:vevhier}
v_{R}, \bar{v}_{R} \gg  v_2, v_1\gg v_1^\prime  = v_2^\prime \approx 0 \qquad\text{and}\qquad
 \alpha_1 = \alpha_2=\alpha_s \approx 0 \ .
\ee
The minimization of the scalar potential yields an $R$-parity violating, or a charge violating, vacuum.  The simplest and most efficient way to avoid either is to introduce one-loop Coleman-Weinberg
effective potential terms, generated by right-chiral leptons coupling to the
$\Delta^c$ field:
\begin{equation}
V_{\rm eff}^{\rm 1-loop}= \frac{1}{16 \pi^2} \sum_i(-1)^{2s} (2s+1)M_i^4 \left
[\ln \left (\frac{M_i^2}{\mu^2}\right)-\frac32 \right ]
\end{equation}
Expanding this potential in the limit in which the SUSY breaking parameters are
small with respect to the triplet VEVs ($v_{R}, \bar v_{R}$), one  obtains an
effective form of the potential in terms of the small parameter:
$$x=\frac{ {\rm Tr}(\Delta^c \Delta^c) {\rm Tr}((\Delta^{c\dagger} \Delta^{c
\dagger})} { [{\rm Tr}(\Delta^{c\dagger} \Delta^c)]^2}.$$
To lowest order in $x$, the effective quadratic term in the one-loop potential becomes:
\begin{eqnarray}
V_{\rm quad.}^{\rm 1-loop}& \simeq &-\frac{|f|^2(M_{\tilde l})_R^2 {\rm Tr}(\Delta^c
\Delta^c){\rm Tr}(\Delta^{c\dagger} \Delta^{c\dagger}) }{128 \pi^2 |v_{R}|^2}
\left \{ (a_1-a_2)g_R^2 \Big (2 \ln\frac{|fv_{R}|^2}{\mu^2}\right.  \nonumber\\
&& \left. \left. +\ln x -2 \ln
2-2 \Big)-[ 2+(a_1+a_2)g_{B-L}^2\right ] \left (\ln x- 2 \ln2 \right )  
\right \}
\label{eq:1st_order}
\end{eqnarray}
Here $a_1$ and $a_2$ correction terms which vanish in the SUSY limit (when $D$-terms vanish) and
$(M_{\tilde l})_R$ are soft right-handed scalar lepton masses.  Before introducing the one-loop corrections, the global minimum contained at
least one doubly-charged Higgs boson with zero or negative mass, but after one-loop corrections
all the masses are positive and the masses become very predictive.

The Higgs boson spectrum of this model was previously analyzed in 
\cite{Frank:2011jia},  which included constraints from FCNC processes from $\epsilon_K$, $K^0-{\bar K}^0$, $D^0-{\bar D}^0$ and $B_{d,s}^0-{\bar B}_{d,s}^0$ data. Here we re-evaluate the masses and mixings to account for the lightest CP-even Higgs boson with a mass of $\sim 125 $ GeV, and to obey restrictions on the spectrum arising from recent constraints on the other Higgs boson masses. We review these below.  

For the doubly charged Higgs bosons, the most up-to-date mass bounds have been obtained  through the direct searches at the LHC.  The ATLAS Collaboration has looked for doubly charged Higgs bosons in pair production of same sign di-lepton final states. Based on the data sample corresponding to an integrated luminosity of 4.7 $\text{fb}^{-1}$ at $\sqrt{s} = 7$ TeV,  masses below 409 GeV, 375 GeV and 398 GeV have been excluded  for $e^{\pm}e^{\pm}$, $e^{\pm}\mu^{\pm}$ and $\mu^{\pm}\mu^{\pm}$, respectively,  assuming a branching ratio of $100\%$ for each final state \cite{ATLAS:2012ai}. 
The CMS Collaboration also searched for the pair production  $pp \rightarrow H^{\pm\pm}H^{\mp \mp}$ and for the associated production 
$pp \rightarrow H^{\pm\pm}H^{\mp}$, in which the masses of $H^{\pm\pm}$ and $H^{\mp}$ are assumed to be degenerate. Using three or more isolated charged lepton final states, the lower limit on $M_{H^{\pm\pm}}$  was found to be between 204 and 459 GeV in the 100\% branching fraction scenarios. Specifically, for $e^\pm e^\pm$, $e^\pm\mu^\pm$,  $e^\pm \tau^\pm$, $\mu^\pm \mu^\pm$,  $\mu^\pm \tau ^\pm$, and $\tau^\pm \tau^\pm$, the 90 \% C.L. limits obtained are  444 GeV, 453 GeV, 373 GeV, 459 GeV, 375 GeV, and 204 GeV respectively  \cite{Chatrchyan:2012ya}. In our work we assume that the decay $H^{\pm \pm} \to \tau^\pm \tau^\pm$ dominates, while the others are negligible, allowing the lower limit of the doubly charged Higgs boson mass to be consistent with LHC searches.   

In LRSUSY the tree-level contribution for the lightest CP-even scalar mass is given by $\frac{1}{2} \sqrt{g_{L}^{2}+g_{R}^{2}}\,v\,|\cos 2\beta|$ \cite{Huitu:1997rr}. Since we do not include left-handed triplets\footnote{Since right-handed triplets do not contribute directly to the
masses of $W_L$ and $Z_L$ bosons, the $\rho$ parameter will be unchanged,
save for a mixing term between $W_L$ and $W_R$ constrained to be
negligibly small.} 
, we can treat $g_{R}$ as a free parameter. If we assume $g_{R}=g_{L}$ (at the electroweak scale) the tree-level mass bound is lifted to $113$~GeV, so the radiative corrections needed for a $125$~GeV Higgs are much smaller than in the MSSM.
The tree-level mass bound has an effect on the allowed range of $\tan \beta$. The radiative corrections depend mostly on stop and sbottom masses. If we fix the third generation squark masses,  there will be a lower bound on the value of $\tan \beta$. For the squark masses in our scans the allowed range is $\tan \beta > 6$. If the squarks are assumed to be heavier, the bound on $\tan \beta$ will become weaker.
\begin{table}[h!]
\begin{ruledtabular}
\begin{tabular}{c c c c}
Fixed Parameters & BP1 & BP2 & BP3 \\
\hline
$M_1$ & 250 & 150 & 250 \\
$M_{2L}$ & 500 & 200 & 500 \\
$M_{2R}$ & 500 & 200 & 500 \\
$M_3$ & 1500 & 1500 & 1500 \\
$v_s$ & 2000 & 2000 & 2000 \\
$Y_{\nu}$ & 0.00003 & 0.00003 & 0.00003 \\
$(M_A)_{min}$ & 300.0 & 300.0 & 300.0\\
$(M_{H^{\pm\pm}})_{min}$ & 200 & 200 & 200\\
\end{tabular}
\caption{Fixed input parameters for the LRSUSY model. The masses are in GeV.}
\label{fixed:param}
\end{ruledtabular}
\end{table}
We show three benchmark points for the parameters of the LRSUSY model. We fix gaugino masses, the singlet VEV, the neutrino Yukawa coupling, 
the doubly charged and pseudoscalar Higgs masses as shown in Table \ref{fixed:param}. The last five parameters mentioned are the same for all benchmarks.

As we wish to study the two-photon decay channel, which is influenced by all charged particles, the benchmark scenarios are designed to make some of the charged particles light. The doubly charged Higgs is light in all benchmarks. In BP2 we have light charginos and in BP3 we have light staus.

In Table \ref{input:param}, we show the parameters on which the scanning is done.
Note that the BP1 and BP3 benchmarks have the same gaugino masses, but are differentiated by running over different soft right-handed slepton masses (heavier in BP1, lighter in BP3 benchmarks).
%%%%%%%%%%%%%%%%%%%
\begin{table}[h!]
\begin{ruledtabular}
\begin{tabular}{c c c c c c c}
Parameter & \multicolumn{2}{c}{BP1} & \multicolumn{2}{c}{BP2} & \multicolumn{2}{c}{BP3} \\ 
~ & Minimum & Maximum & Minimum & Maximum & Minimum & Maximum\\
\hline
tan$\beta$ & 2.5 & 40 & 2.5 & 40 & 2.5 & 40\\
$\lambda$ & 0.3 & 0.6 & 0.3 & 0.6 & 0.3 & 0.6\\
$(M_{\tilde q})_{L}$(GeV) & 1000 & 1500 & 1000 & 1500 & 1000 & 1500\\
$(M_{\tilde q})_{R}$(GeV) & 1000 & 1500 & 1000 & 1500 & 1000 & 1500\\
$(M_{\tilde \ell})_{L}$(GeV) & 1000 & 1500 & 1000 & 1500 & 100 & 300\\
$(M_{\tilde \ell})_{R}$(GeV) & 1000 & 1500 & 1000 & 1500 & 500 & 700\\ 
f & 0.4 & 0.7 & 0.4 & 0.7 & 0.55 & 0.7\\
$M_R$(GeV) & 100000 & 150000 & 100000 & 150000 & 100000 & 150000\\
A & -2000 & 2000 & -2000 & 2000 & -2000 & 2000 \\
$\mu_{eff}$(GeV) & -700 & -400 & -500 & -200 & -700 & -400\\
$v_R$(GeV) & 4000 & 5000 & 4000 & 5000 & 3000 & 4500\\
tan$\delta$ & 0.93 & 0.99 & 0.93 & 0.99 & 0.96 & 0.99\\ 
\end{tabular}
\caption{Ranges of the input parameters in the scan used to compute the Higgs production and decays for the BP1, BP2 and BP3 benchmarks.}
\label{input:param}
\end{ruledtabular}
\end{table}
%%%%%%%%%%%%%

\begin{figure}[b!]
\begin{center}
\includegraphics[height=2.0in,width=1.9in]{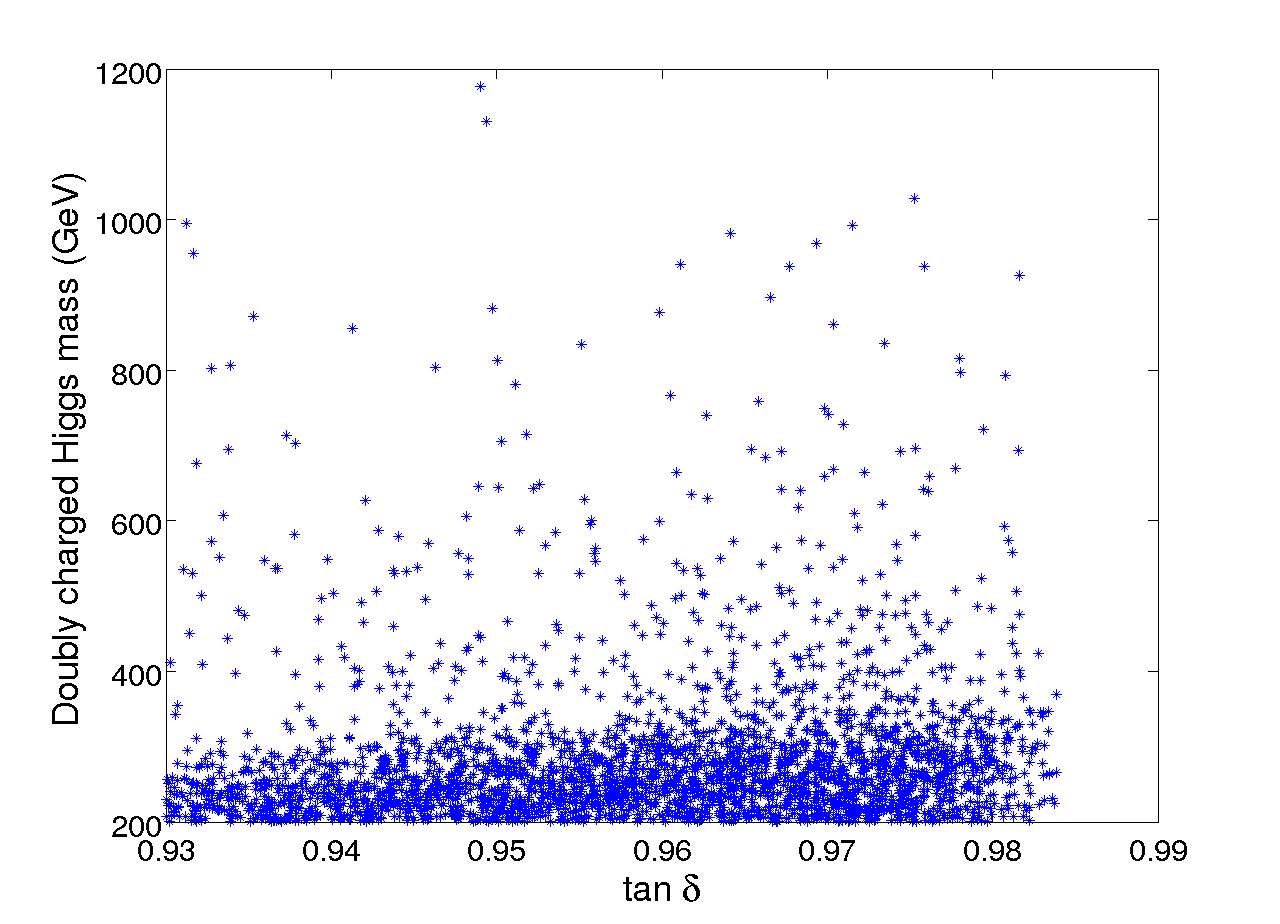}
\includegraphics[height=2.0in,width=1.9in]{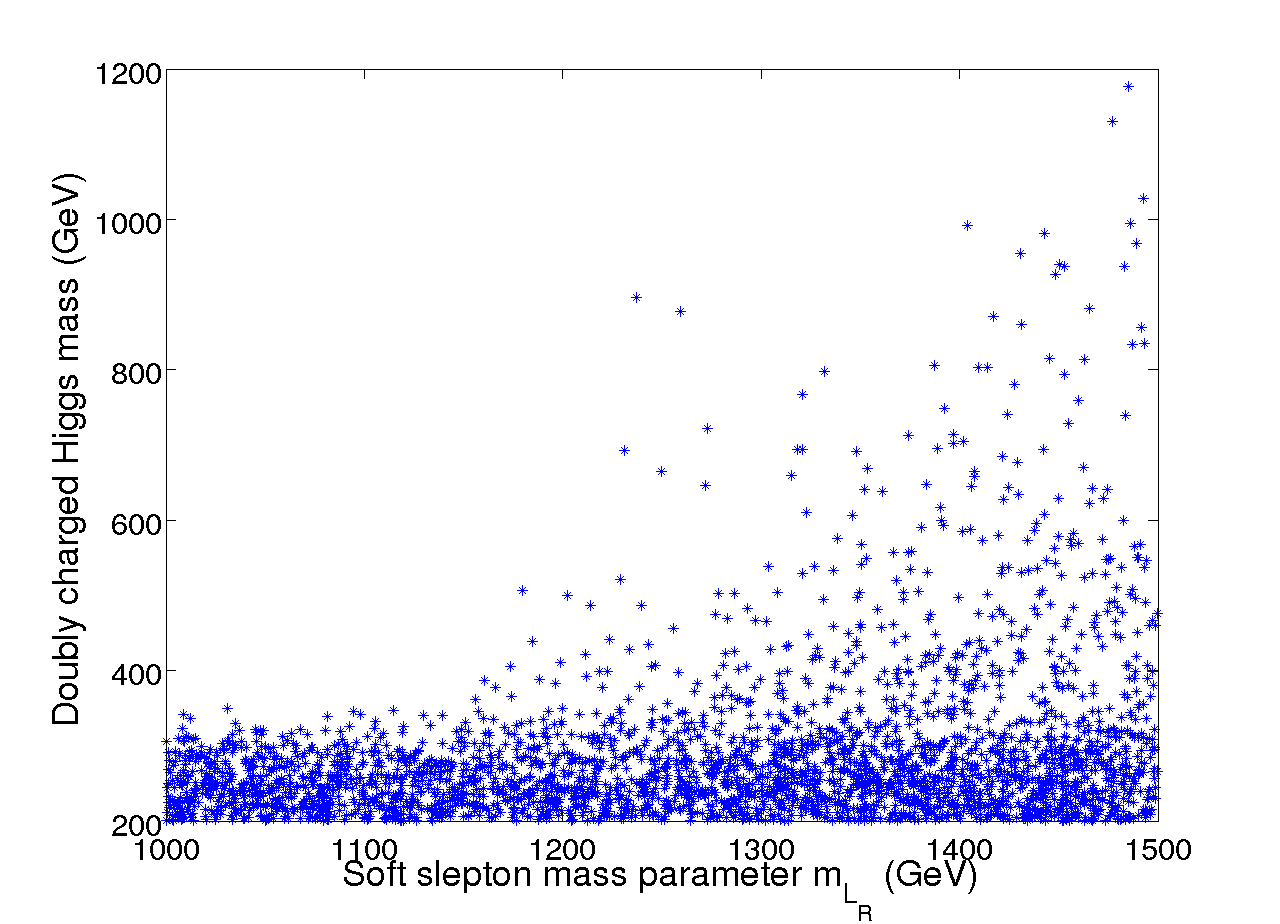}
\includegraphics[height=2.0in,width=1.9in]{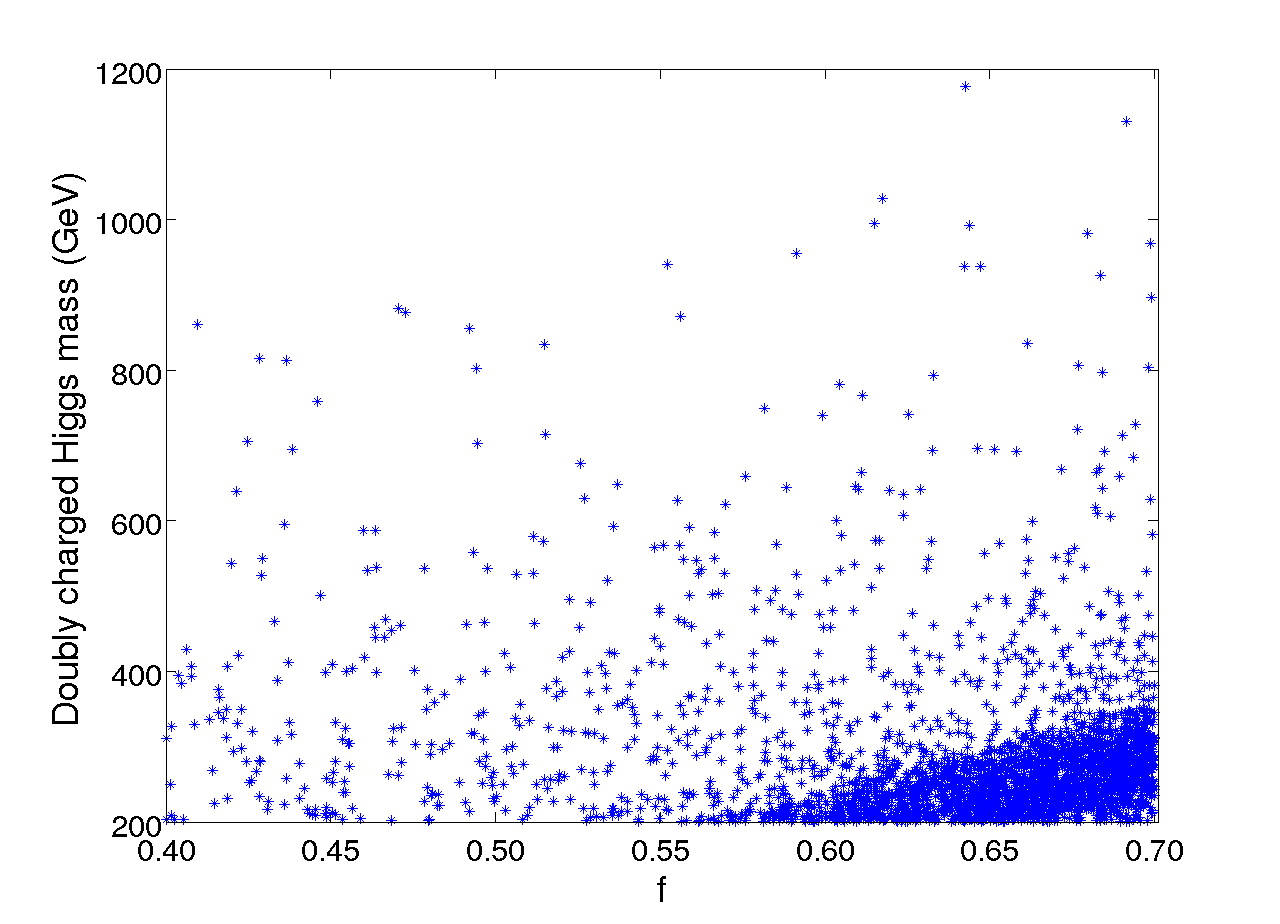}
\caption{Dependence of the mass of the doubly charged Higgs boson on the parameters of the model, as given in the Table \ref{input:param}. We show the variation with  $\tan \delta$, the soft mass for the right-handed $m_{L_R}$, and the triplet Higgs Yukawa coupling, $f$.} 
 \label{fig:mHpmpm}
\end{center}
\end{figure}

\begin{figure}[t!] 
\begin{center}
$\begin{array}{cc}
\hspace*{-1.5cm}
\includegraphics[width=3.0in,height=2.5in]{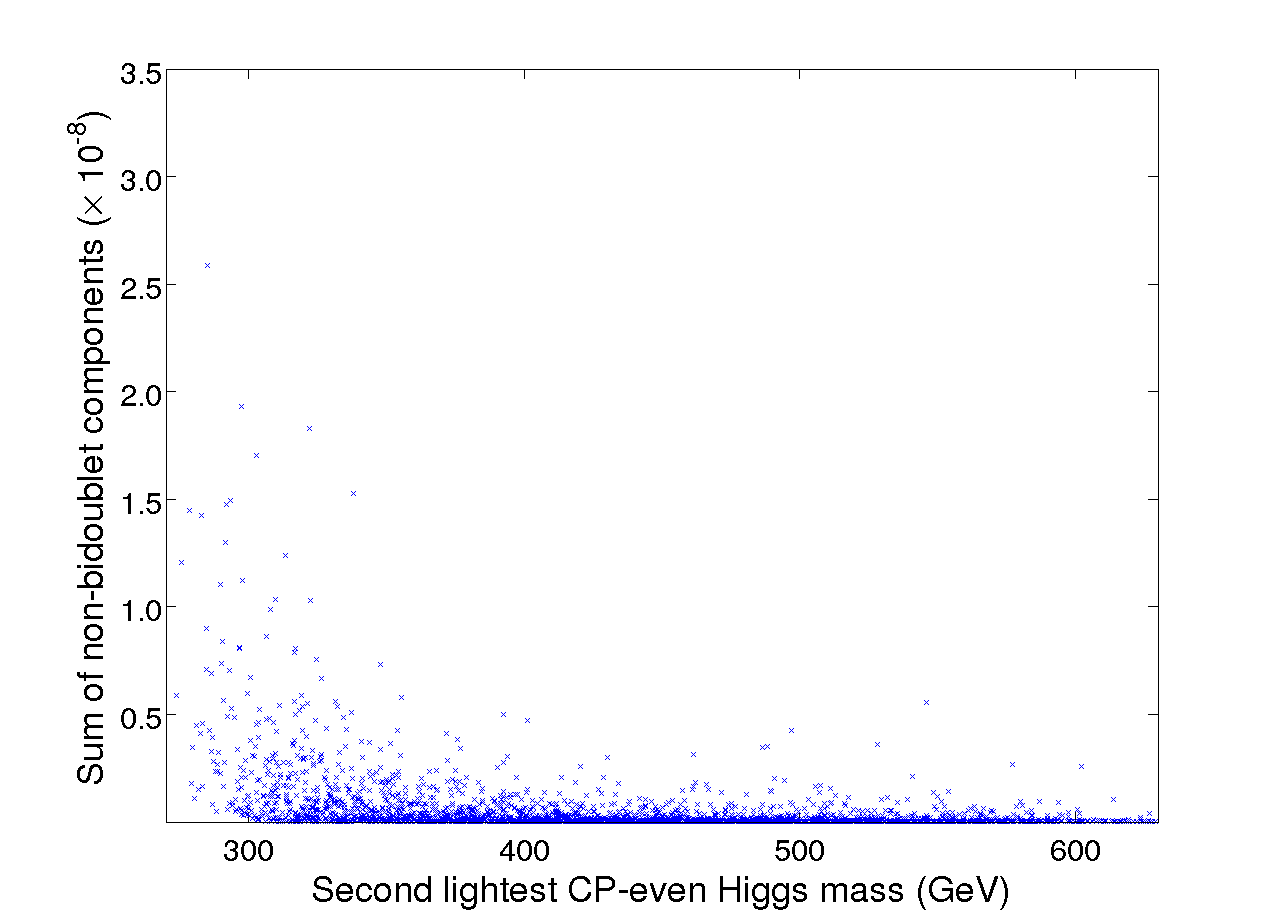} &
\includegraphics[width=3.0in,height=2.5in]{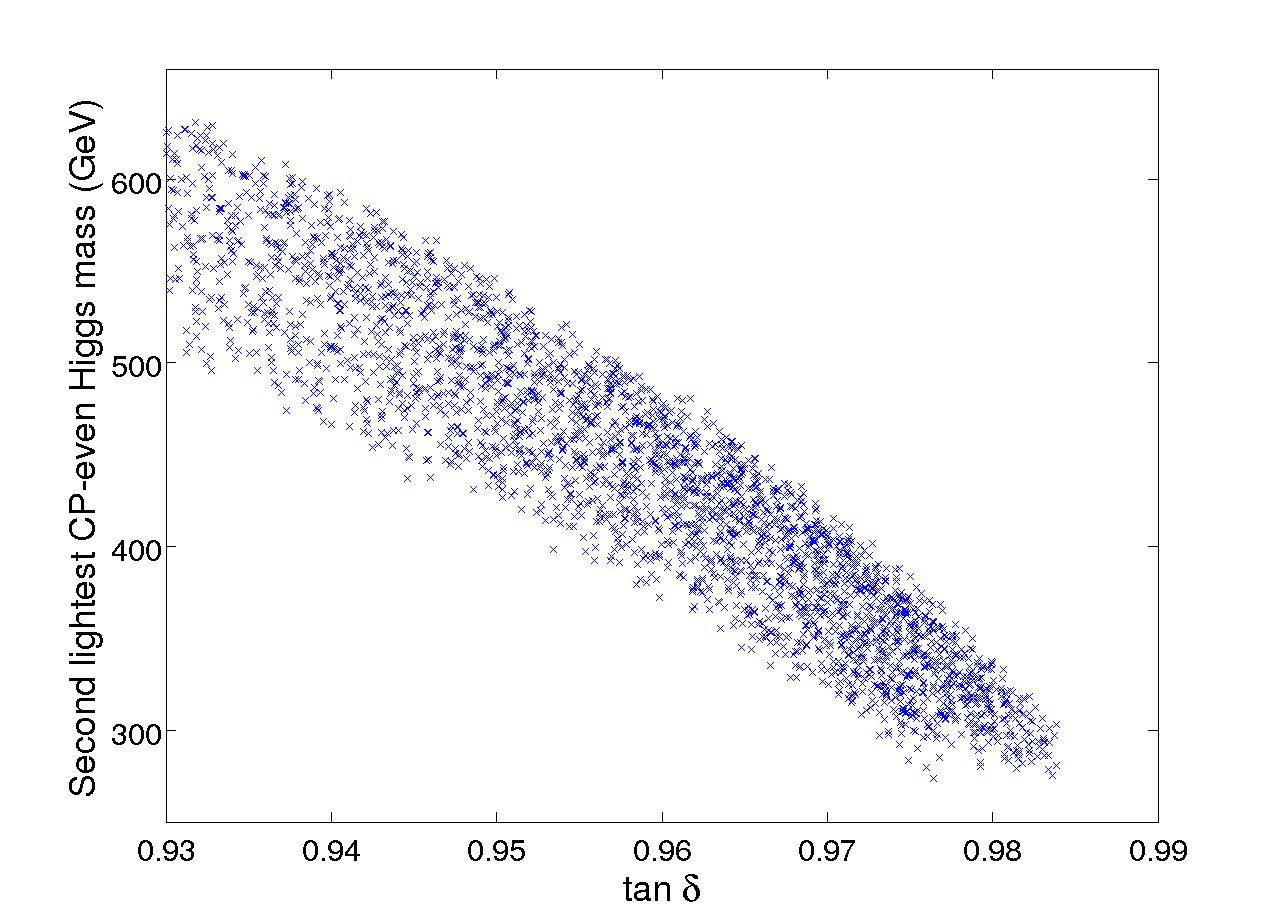} \\
\hspace*{-1.5cm}
\includegraphics[width=3.0in,height=2.5in]{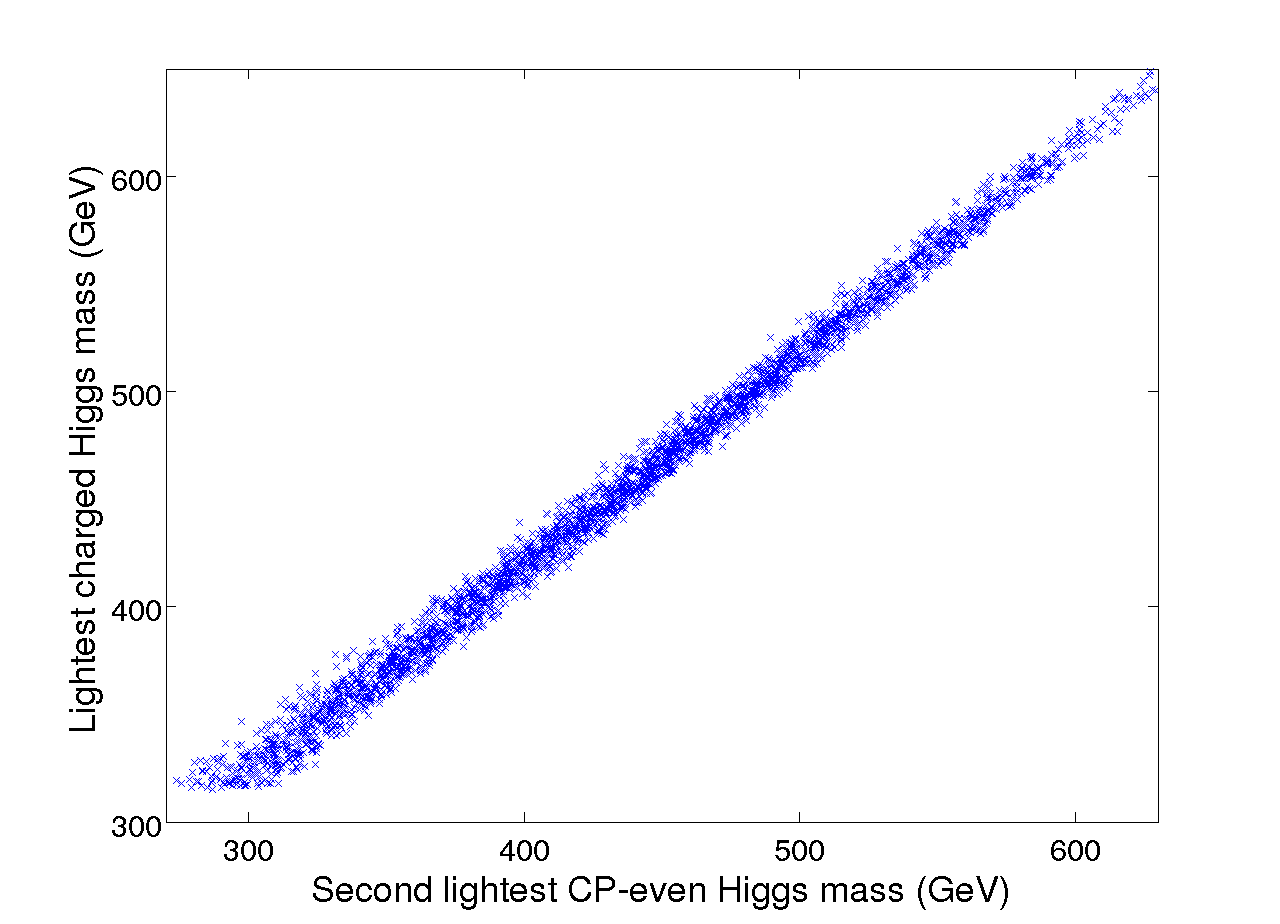} & 
\includegraphics[width=3.0in,height=2.5in]{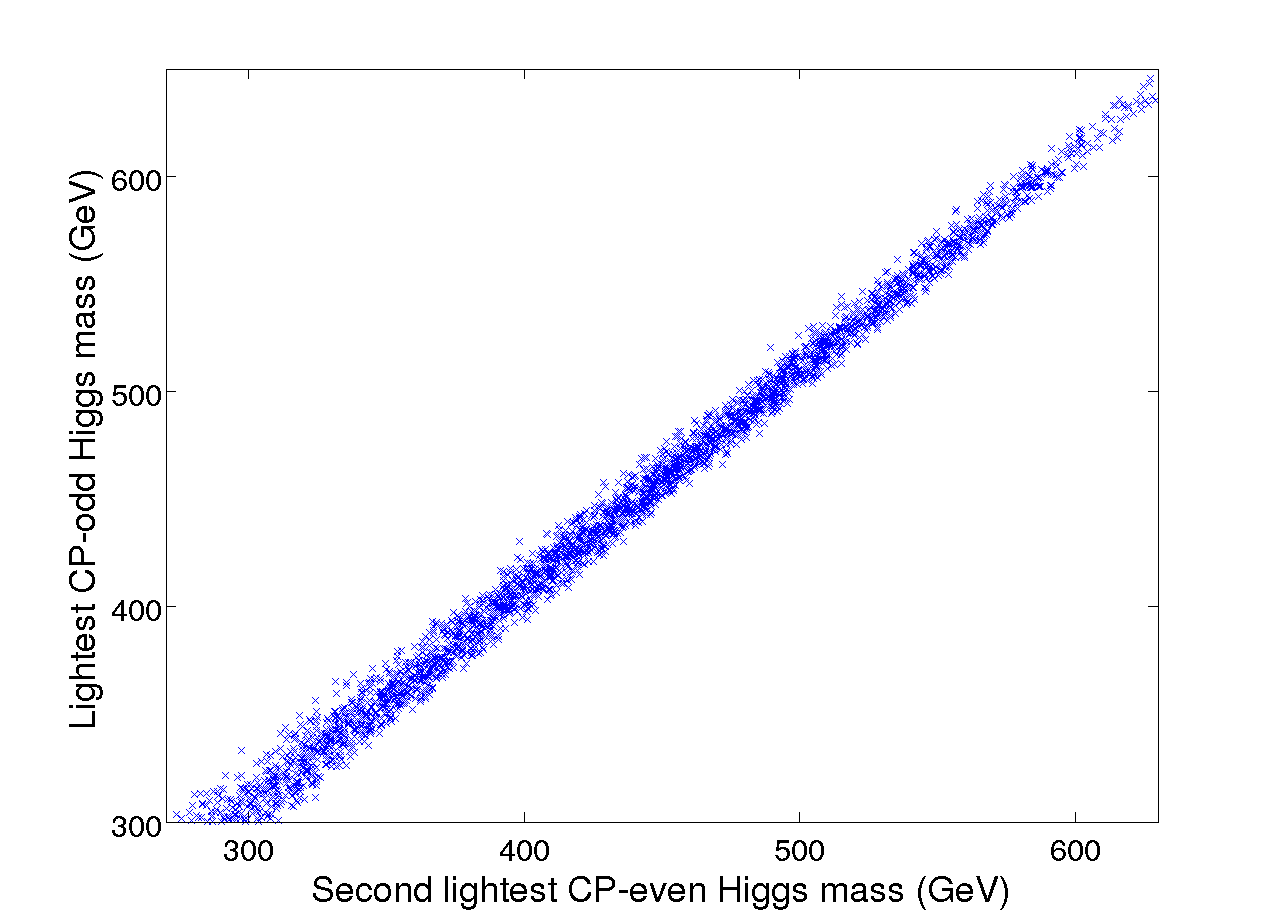}
\end{array}
$ 
\end{center}
\vskip -0.5cm
\caption{Dependence of the mass of the second lightest CP-even Higgs boson on the parameters of the model, as given in the Table \ref{input:param}. We show the triplet composition of the state (upper left panel), the variation with  $\tan \delta$ (upper right panel), dependence on the lightest singly charged Higgs mass $M_{H^\pm}$ (lower left panel) and on the lightest pseudoscalar mass $M_A$ (lower right panel).} 
\label{fig:mH}
\end{figure}

The lightest CP-even state in our model (with mass $M_{H^0_1}\sim 125 $ GeV) is SM-like, that is, it is composed mainly of Higgs bidoublet components.  The next lightest state is the doubly charged Higgs boson $H_1^{\pm \pm}$. Its mass varies in the 200-300 GeV region, and is sensitive to the ratio of the two triplet Higgs VEVs,  $\tan \delta=\frac {\bar{v}_{R}}{v_{R}}$, the soft mass for the right-handed sleptons $(M_{\tilde l})_{R}$, and the triplet Higgs Yukawa coupling, $f$. We show these variations in the plots in Fig. \ref{fig:mHpmpm}.  We note that no masses for the doubly charged Higgs boson are obtained if $\tan \delta$ increases beyond 0.983. The values of $\tan \delta$  are restricted by  the experimental bounds for the lightest CP-odd scalar and  
the charged Higgs masses, as shown in Fig. \ref{fig:mH}. If the limit on  
these masses is relaxed, points satisfying the  
doubly charged mass limit can be obtained with $\tan \delta$ values close to one. The scans are performed in the range $0.93<\tan\delta<0.99$. At the lower end it becomes more difficult to satisfy the bound on the doubly charged Higgs mass. This is due to the fact that the diagonal elements of the doubly charged Higgs mass matrix have terms $\pm 2g_{R}^{2}(v_R^2-\bar{v}_R^2)$ \cite{Frank:2011jia}. Hence if $\tan \delta$ deviates largely from one, one of these values will become increasingly negative and hence larger radiative corrections are needed to satisfy the bound on the doubly charged Higgs mass. With the range we use for $v_R$, the doubly charged Higgs mass constraint requires $\tan \delta > 0.9$ and there are very few points which survive the $200$~GeV constraint in the lower end of the range. On the other hand it is not possible to have smaller values of $v_R$ since that would result in a too low a $W_R$ mass.

The doubly charged Higgs mass depends on the soft right-handed slepton mass, as in Eq. \ref{eq:1st_order}, since this parameter determines the amount of supersymmetry breaking in the $\tau-{\tilde \tau}$ loops.
 However, masses in the 200-300 GeV region are obtained for all values $(M_{\tilde l})_{R}$ in the 1-1.5 TeV region; while heavier masses are more likely for heavier slepton masses. The most striking impact on the doubly charged Higgs mass comes from  
the parameter $f$, the Yukawa coupling for the triplet Higgs bosons, which is the coefficient of the term that gives radiative corrections from $\tau-{\tilde \tau}$ loops to the doubly charged Higgs mass. While parameter points with $f \in (0.4, 0.6)$ range exist, the experimental constraints  on the doubly charged Higgs mass clearly favor large triplet Yukawa couplings, $f >0.6$ for soft slepton masses around $1$~TeV. For larger soft slepton masses one can get viable doubly charged Higgs masses with smaller values of $f$ also.
The other radiative contributions to the doubly charged Higgs mass come from the singlet Higgs loops, but that contribution is large only if we let the ef\mbox{}fective $\mu$-parameter be in the TeV range. This will lead to fine tuning in the Z boson mass.
The second lightest CP-even Higgs boson is predicted to be relatively light\footnote{This is the second FCNC-conserving neutral Higgs boson.}. This Higgs boson is also mostly doublet-like: we show its composition in the left-hand upper panel of Fig. \ref{fig:mH}, where it is seen that the triplet components are negligible. The dependence of its mass  with $\tan \delta$ (upper right-hand panel), with the mass of the lightest singly charged Higgs boson (lower left-hand panel), and with the mass of the lightest pseudoscalar (lower right-hand panel) is almost linear. A heavier neutral Higgs favors $\tan \delta \sim 0.93$, while the mass splittings between the second lightest CP-even neutral Higgs and the lightest singly charged Higgs, as well as  the pseudoscalar,  are $M_{H_2^0}-M_{H_1^\pm} \sim M_{H_2^0}-M_{A} \lesssim 50$ GeV.
\begin{table}[h!]
\begin{ruledtabular}
\begin{tabular}{c c c c c c c}
Component & $H_1^0$ & $H_2^0$ & $H_3^0$ & Component & $H_1^+$ & $H_1^{++}$\\
\hline
$\phi_1^0$ & $99.65\%$ & $0.01\%$ & --- & $\phi_1^+$ & $99.58\%$ & \\
$\phi_2^0$ & $0.01\%$ & $99.65\%$ & --- & $\chi_1^+$ & $0.34\%$ & \\
$\chi_1^0$ & --- & $0.34\%$ & --- & $\delta^{c+}$ & $0.04\%$ & \\
$\chi_2^0$ & $0.34\%$ & --- & --- & $\bar\delta^{c+}$ & $0.04\%$ & \\
$\delta^{c0}$ & --- & --- & $52.10\%$ & $\delta^{c++}$ &  & $52.11\%$\\
$\bar\delta^{c0}$ & --- & --- & $47.90\%$ & $\bar\delta^{c++}$ & & $47.89\%$\\
\hline
Mass (GeV) & $123.5$ & $436.3$ & $1988$ & & $455.4$ & $286.8$\\
\end{tabular}
\caption{The average compositions and masses of the lightest Higgs bosons for benchmark BP1. For benchmark BP2 the results are similar except for the mass of $H_3^0$, which is $2280$~GeV. For benchmark BP3 the compositions are similar but the masses are somewhat lighter: $351.0$~GeV ($H_2^0$), $1818$~GeV ($H_3^0$), $370.3$~GeV ($H_1^+$) and $238.7$~GeV ($H_1^{++}$).\label{tb:higgsmass}}
\end{ruledtabular}
\end{table}
The average compositions and masses of the lightest Higgs bosons are shown in Table \ref{tb:higgsmass}.
%

%\newpage
%%%%%%%%%%%%%%%%%%%%%%%%%%%%%%%%%%%%%%
\section{The Supersymmetric Spectrum}\label{sec:cspectrum}
%%%%%%%%%%%%%%%%%%%%%%%%%%%%%%%%%%%%
First, we assume that squarks and gluinos are heavy, in agreement with the LHC 
limits. Direct searches for squarks and gluinos require their soft mass terms to be at least at TeV scale. Squark masses below $780$ GeV and gluino masses of up to $1.1$--$1.2$ TeV are excluded at 95\% CL within  several models, for LSP masses below 100 GeV at CMS \cite{Chatrchyan:2014lfa}. The third generation squark masses need to be large also to generate the radiative corrections to the Higgs mass. In contrast,  neutralinos \cite{Abdallah:2003xe}, charginos and sleptons \cite{Chatrchyan:SUS13006} are still allowed to be lighter.

The mass spectrum of the lightest superpartners is largely determined by the soft supersymmetry breaking parameters. Since they are in principle unknown, there are for instance several viable alternatives for the LSP. The different options for dark matter in LRSUSY have been studied in \cite{Demir:2006ef,Esteves:2011gk}. 
%%%%
\subsection{Charginos}\label{subsec:charginos}
%%%%%%%%%%%%%%%%%%%%%%%%%%%%%%%%%%%%%%%%%%%%%%%%%%%%%%%%%%%%%%%%%%%%%%%%%%
In this model there are five singly charged charginos, which can be given in 
the following basis $(\tilde{\Delta}^{\pm}, \tilde{\Phi}_{1}^{\pm},\tilde{\Phi}_{2}^{\pm}, \W_{L}^{\pm}, \W_{R}^{\pm})$. The mass matrix is

\begin{equation}
\label{eq:mchargino}
M_{\tchi^{\pm}}=
\bpm
\lambda v_{s}/\sqrt{2} & 0 & 0 & 0 & -g_{R}v_{R}\\
0 & 0 & \mu_{\mathrm{eff}} & g_{L}v_{u}/\sqrt{2} & 0\\
0 & \mu_{\mathrm{eff}} & 0 & 0 & -g_{R}v_{d}/\sqrt{2}\\
0 & 0 & g_{L}v_{d}/\sqrt{2} & M_{2L} & 0\\
g_{R}\overline{v}_{R} & -g_{R}v_{u}/\sqrt{2} & 0 & 0 & M_{2R}\\
\epm
\end{equation}

The mass spectrum essentially depends on $M_{2L}$, $M_{2R}$, $\mu_{\mathrm{eff}}$, $v_{R}$ and $\overline{v}_{R}$ (or $\tan \delta$).
The masses are close to these parameter values with corrections of few tens of GeV's.

The lightest chargino has a mass slightly below the soft gaugino masses and it is mostly a mixture of a bidoublet 
higgsino and a left-handed wino. One chargino is essentially a pure bidoublet higgsino with a mass $|\mu_{\mathrm{eff}}|$. 
There is also a third sub-TeV chargino, which is also a combination of a bidoublet Higgsino and left-handed wino.
The two heaviest charginos have masses of the order of $v_{R}$, i.e. they are in the multi-TeV region. 
These are mostly composed of the right-handed wino and the singly charged $\delta^{c-}\,(\bar \delta^{c+})$.

The lightest chargino can always decay via $\tchi^{+}_{1}\to W^{+}
\tchi^{0}_{1}$. The decay to $W^{+}\tchi_{2}^{0}$ is kinematically forbidden. 
In the case BP3 also $\tilde{\tau}\nu$ and $\tau\tilde{\nu}$ can be possible 
decay channels. Within the parameter regions the lighter stau could be lighter than 
the lightest neutralino. We discard such points. 

The doubly charged higgsino has a mass $|\lambda v_{s}/\sqrt{2}|$, and therefore can be light and can have interesting collider 
signatures \cite{Huitu:1993gf,Chacko:1997cm,Raidal:1998vi,Demir:2009nq,Demir:2008wt,Frank:2007nv,Alloul:2013raa}. 
%as can be seen from the chargino/neutralino Lagrangian.  
Since $\lambda$ and $v_{s}$ are in principle unconstrained, the mass of the doubly charged higgsino can vary over a wide region. Since we assume that the doubly charged Higgs coupling to taus is large to ensure large radiative corrections to the doubly charged Higgs mass and  also its decay to $\tau^\pm\tau^\pm$ to dominate, the doubly charged higgsino will decay to $\tau^{\pm}\tilde{\tau}^{\pm}$ unless that mode is kinematically forbidden. If the stau is heavy, $\tchi^{++}\to H^{++}_{1}\tchi_{1}^{0}$ will be the dominant decay mode \cite{Babu:2013ega}. For our benchmarks $v_s \sim $ TeV, and the doubly charged higgsinos are heavy.
%%%%%%%%%
\subsection{Neutralinos}\label{subsec:neutralinos}
%%%%%%%%%%%%%%%%%%%%%%%%%%%%%%%%%%%%%%%%%%%%%%%%%%%%%%%%%%%%%%%%%%%%%%%%%
There are ten neutralinos in the spectrum. The neutralino mass matrix in the basis $(\tilde{\phi}_{1},\tilde{\phi}_{2}, \tilde{\chi}_{1}, \tilde{\chi}_{2},\tilde{\delta}^{c},\overline{\tilde{\delta}^{c}},\tilde{S},\B,\W_{L}^{0},\W_{R}^{0})$ is

\begin{align}
 &  M_{\tchi^{0}}  = \nonumber \\
 & \bpm
0 & 0 & 0 & -\mu_{eff} & 0 & 0 & -\mu_{d} & 0 & g_{L}v_{u}/\sqrt{2} & -g_{R}v_{u}/\sqrt{2}\\
0 & 0 & -\mu_{eff} & 0 & 0 & 0 & 0 & 0 & 0 & 0\\
0 & -\mu_{eff} & 0 & 0 & 0 & 0 & 0 & 0 & 0 & 0\\
-\mu_{eff} & 0 & 0 & 0 & 0 & 0 & -\mu_{u} & 0 & -g_{L}v_{d}/\sqrt{2} & g_{R}v_{d}/\sqrt{2}\\
0 & 0 & 0 & 0 & 0 & \lambda v_{s}/\sqrt{2} & \lambda \overline{v}_{R}/\sqrt{2} & g'v_{R} & 0 & -g_{R}v_{R}\\
0 & 0 & 0 & 0 & \lambda v_{s}/\sqrt{2} & 0 & \lambda v_{R}/\sqrt{2} & -g'\overline{v}_{R} & 0 & -g_{R}\overline{v}_{R}\\
-\mu_{d} & 0 & 0 & -\mu_{u} & \lambda \overline{v}_{R}/\sqrt{2} & \lambda v_{R}/\sqrt{2} & 0 & 0 & 0 & 0\\
0 & 0 & 0 & 0 & g'v_{R} & -g'\overline{v}_{R} & 0 & M_{1} & 0 & 0\\
g_{L}v_{u}/\sqrt{2} & 0 & 0 & -g_{L}v_{d}/\sqrt{2} & 0 & 0 & 0 & 0 & M_{2L} & 0\\
-g_{R}v_{u}/\sqrt{2} & 0 & 0 & g_{R}v_{d}/\sqrt{2} & -g_{R}v_{R} & -g_{R}\overline{v}_{R} & 0 & 0 & 0 & M_{2R}
\epm, %\nonumber
\end{align}
where $\mu_{u,d}=\lambda_{12} v_{u,d}/\sqrt{2}$.

With the superpotential and the parameter ranges used, the lightest neutralino is dominantly a singlino with a relatively large $\W^{0}_{R}$ component. However the composition of the lightest neutralino depends crucially on the form of the superpotential. Terms of the form $M_{S}S^{2}$ and $\lambda_{S}S^{3}$ are gauge invariant and could be added to the superpotential if there are no additional symmetries that would forbid their existence. These terms contribute to the mass of the singlino-dominated state and may easily make it heavier than the lightest gaugino-dominated state.

The lightest gaugino-dominated state is mostly $\W^{0}_{L}$. Since the bino mixes with the neutral components of the fermionic triplets $\tilde{\Delta}^{c}$ and $\tilde{\bar{\Delta}}^{c}$, and the mixing terms $\B\tilde{\Delta}^{c}$ and $\B\tilde{\bar{\Delta}}^{c}$ are proportional to $v_{R}$ and $\bar{v}_{R}$, respectively, even if the bino mass $M_1$ is the lightest gaugino mass, through mixing the bino dominated states will be heavier.  
%The $\W^{0}_{L}$ state has a mass  slightly below $|M_{2L}|$.  
If $|M_{2L}|<|\mu_{\mathrm{eff}}|$, the $\W^{0}_{L}$ is the second lightest neutralino (with mass 
slightly below $|M_{2L}|$), otherwise the  
 bidoublet higgsino would be second lightest, with a mass $\sim|\mu_{\mathrm{eff}}|$.

For the case BP1 the average compositions of the lightest neutralinos and charginos are given in Table \ref{tb:charneutr1}. The results for the BP3 case are essentially the same.
\begin{table}[h]
\begin{ruledtabular}
\begin{tabular}{c c c c c c c}
Component & $\tchi_1^0$ & $\tchi_2^0$ & $\tchi_3^0$ & Component & $\tchi_1^+$ & $\tchi_2^+$\\
\hline
$\tilde{\phi}_1$ & $0.84\%$ & $16.44\%$ & --- & $\tilde{\phi}^+$ & --- & $71.98\%$\\
$\tilde{\phi}_2$ & --- & --- & $50.00\%$ & $\tchi^+$ & $52.04\%$ & ---\\
$\tchi_1$ & --- & --- & $50.00\%$ & $\tilde{\delta}^c+$ & --- & $3.07\%$\\
$\tchi_2$ & $0.29\%$ & $24.25\%$ & --- & $\W_L^+$ & $47.96\%$ & ---\\
$\tilde{\delta}^{c}$ & $0.17\%$ & --- & --- & $\W_R^+$ & --- & $24.96\%$\\
$\tilde {\bar \delta}^c$ & $0.19\%$ & --- & --- & & &\\
$\tilde{S}$ & $69.38\%$ & $0.09\%$ & --- & & &\\
$\B$ & $0.06\%$ & --- & --- & & &\\
$\W_{L}$ & $0.09\%$ & $59.17\%$ & --- & & &\\
$\W_{R}$ & $28.97\%$ & $0.03\%$ & --- & & &\\
\hline
Mass (GeV) & $152.4$ & $438.8$ & $548.0$ & & $456.1$ & $547.6$\\
\end{tabular}
\end{ruledtabular}
\caption{The average masses and compositions of the lightest neutralinos and charginos for  benchmark BP1. The values are almost the same for benchmark BP3.\label{tb:charneutr1}}
\end{table}
The average masses and compositions (for BP2) of the lightest neutralinos and charginos are given in Table \ref{tb:charneutr2}. With these parameters the charginos and neutralinos are lighter and the bidoublet states decouple and do not mix with the triplets or right-handed gaugino states as much as in BP1 or BP3 cases.

\begin{table}[h]
\begin{ruledtabular}
\begin{tabular}{c c c c c c c}
Component & $\tchi_1^0$ & $\tchi_2^0$ & $\tchi_3^0$ & Component & $\tchi_1^+$ & $\tchi_2^+$\\
\hline
$\tilde{\phi}_1$ & $0.60\%$ & $5.23\%$ & --- & $\tilde{\phi}^+$ & --- & $99.83\%$\\
$\tilde{\phi}_2$ & --- & --- & $50.00\%$ & $\tchi^+$ & $54.91\%$ & ---\\
$\tchi_1$ & --- & --- & $50.00\%$ & $\tilde{\delta}^c+$ & --- & $0.16\%$\\
$\tchi_2$ & $1.42\%$ & $18.17\%$ & --- & $\W_L^+$ & $45.09\%$ & ---\\
$\tilde{\delta}^{c}$ & $0.03\%$ & --- & --- & $\W_R^+$ & --- & $0.01\%$\\
$\tilde {\bar \delta}^c$ & $0.03\%$ & --- & --- & & &\\
$\tilde{S}$ & $63.10\%$ & $0.24\%$ & --- & & &\\
$\B$ & $0.07\%$ & --- & --- & & &\\
$\W_{L}$ & $0.58\%$ & $76.22\%$ & --- & & &\\
$\W_{R}$ & $34.16\%$ & $0.13\%$ & --- & & &\\
\hline
Mass (GeV) & $77.4$ & $174.1$ & $349.2$ & & $184.7$ & $349.0$\\
\end{tabular}
\end{ruledtabular}
\caption{The average masses and compositions of the lightest neutralinos and charginos for benchmark BP2. \label{tb:charneutr2}}
\end{table}

With our  parameter choices for BP1 and BP3 benchmarks, the second lightest neutralino is more than 200 GeV heavier than the lightest one. This means that both of the channels $\tchi^{0}_{2}\to \tchi^{0}_{1}Z$ and $\tchi^{0}_{2}\to \tchi^{0}_{1}h$ are kinematically open. In the case of benchmark BP3 the channels $\tau\tilde{\tau}$ and $\nu\tilde{\nu}$ may be open if the stau or sneutrino are light enough. The $\tilde W^{0}_{L}$-dominated state will dominantly decay to these channels if it is not the NLSP since the largest components in $\tchi^{0}_{1}$ do not couple to $W^{0}_{L}$. In the case BP2 both the lightest and the second lightest neutralino masses are smaller than in the benchmarks BP1 or BP3. In this case the mass splitting is often around $100$~GeV so that $\tchi^{0}_{2}\to \tchi^{0}_{1}h$ is not allowed on-shell. 
 when that occurs, the dominant decays from NLSP to LSP would be the three-body decays $\tchi^{0}_{2}\to \tchi^{0}_{1} f \bar{f}$. The different neutralinos and charginos production in this model can give interesting 
signals at the LHC \cite{Alloul:2013fra}.

\subsection{Sleptons}\label{subsec:sleptons}
%%%%%%%%%%%%%%%%%%%%%%%%%%%%%%%%%%%%%%%%%%%%%%%%%%%%%%%%%%%%%%%%%%%%%%%%
The slepton masses depend on the benchmark chosen. While the right-handed ones need to be heavier to generate a large doubly charged Higgs mass, the left-handed sleptons can be light in the BP3 scenario, as in Table \ref{input:param}. The sneutrino masses are largely determined by the soft slepton masses. For 
benchmarks BP1 and BP2 sneutrinos are heavy, whereas in BP3 case the lightest sneutrino mass 
typically varies between (350 - 700) GeV. Hence for benchmark BP3 the left-handed 
sneutrino may be the second lightest neutral superpartner. If the sneutrino 
is the NLSP it will dominantly decays to $\nu_{L}\tchi^{0}_{1}$,  which 
is invisible.

%
%%%%%%%%%%%%%%%%%%%%%%%%%%%%%%%%%%%%
\section{The decay modes of $H^0_1 $ }
\label{sec:gamgam}

In this section we discuss the light SM like Higgs boson, $H^0_1$ decays 
into the SM final states: $ f {\bar f}, WW,ZZ$, $\gamma \gamma $ and 
gg, where last two decay modes are at one loop level. As evident from the discussion 
in Sec. \ref{sec:model}, the observed scalar at the LHC would have to be
a superposed state of the many physical scalar degrees of freedom in the LRSUSY
Higgs sector. Thus the decay properties of such a Higgs with mass at $\sim$ 125 GeV
would crucially depend on its composition, which in turn would give us an insight
on the parameter space of the model which allows the Higgs to behave as 
the one observed at
the LHC. Note that the partial decay widths for channels which would be affected directly 
by new particles in the spectrum and that can couple to the Higgs boson are the 
loop induced decay modes, namely $H^0_1 \to g~g,~\gamma \gamma,~Z \gamma$. 
The $gg$ mode will be only affected through colored particles appearing in the 
loop. As current limits on the supersymmetric colored states from direct searches at the
LHC are quite strong, they would be quite heavy and therefore should not affect the $H^0_1 \to g~g$
partial width. This is also reflected in the parameter choice that we assume. However, the 
$\gamma \gamma,~Z \gamma$ modes are definitely affected by the particles unique to the
LRSUSY particle spectrum, such as the charged scalars (singly/doubly) 
and fermions and will play a major role.
Note that a somewhat slight discrepancy in the Higgs observation is seen in the $\gamma \gamma$ mode, though reduced in the new CMS data \cite{Khachatryan:2014ira}, 
 makes the aforementioned contributions in this model all the more worth 
considering.   

\subsection{$H^0_1 \to \gamma \gamma $}
The $H^0_1 \to \gamma \gamma $ decay is a loop process involving the 
exchange of spin $0, 1/2, 1$ particles in the loop. In our case, 
in addition to the SM contributions (mainly coming from top and $W$
boson loop) we add contributions from charginos (lightest and second 
lightest states), both lighter and heavier staus, stops, sbottoms, 
singly and doubly charged Higgs bosons. The 
introduction of doubly charged Higgs boson is very crucial as 
this can lead to non-trivial contribution to this partial width
simply because of its enhanced electromagnetic strength.
The most general expression for $\Gamma (H \to \gamma \gamma )$ \cite{Shifman:1979eb, Gunion:1989we, Djouadi:2005gj} in the 
presence of spin $0, 1/2 $ and $1$ particles is given by:
\begin{eqnarray}
\Gamma(H^0_1 \rightarrow\gamma\gamma)
& = & \frac{\alpha^2 M_{H^0_1}^3}
{1024\pi^3}~ \bigg| \sum_f \frac{2 N_c Q_f^2} {m_f}g_{H^0_1ff}  %g_{h f\bar{f}} 
A^h_{1/2}
(\tau_f) + \frac{g_{H^0_1VV}}{M_W^2}A^h_1 (\tau_V) \nonumber \\
&&  - \frac{N_{c,S}Q_S^2}{M_S^2} g_{H^0_1 SS}
A^h_0(\tau_S) \bigg|^2 \, 
\end{eqnarray}
where, $f,V,S$ stands for fermions $(t,b, \tau, \tilde \chi^\pm_1, 
\tilde \chi^\pm_2)$, vectors $(W^\pm )$ and scalars 
$(H^\pm_1, H^{\pm \pm},\tilde t_i, \tilde b_i$, $\tilde \tau_i, i =1,2 )$ 
respectively. Here $g_{H^0_1 f f}\, ,g_{H^0_1 VV}$ and $g_{H^0_1SS}$ 
represent the Higgs couplings with fermions ($f$), 
SM gauge bosons $(W^\pm)$ and 
with scalars ($S$) respectively,  $\alpha$ is the fine-structure 
constant, $N_{c,S}=3 (1)$ for quarks, squarks (leptons, sleptons,
singly and doubly charged scalars), $Q_{f,S}$ and $m_f,M_S$ are 
the electric charge and mass of the fermions and scalars respectively 
in the loop. $\tau_i=M_{H^0_1}^2/4M_i^2~$, 
where $i$ represent fermion, vector and scalar particles as mentioned above. 

The relevant loop functions are given by
\begin{eqnarray}
A_{0}(\tau) &=& -[\tau -f(\tau)] \tau^{-2} \, ,
\label{eq:Ascalar}\\
A_{1/2}(\tau)&=& 2\left[\tau+(\tau-1)f(\tau)\right]\tau^{-2}, 
\label{eq:Afermion}\\
A_1(\tau)&=& -\left[2\tau^2+3\tau+3(2\tau-1)f(\tau)\right]\tau^{-2}, 
\label{eq:Avector} 
\end{eqnarray}
 and the function $f(\tau)$ is given by
\begin{eqnarray}
f(\tau)=\left\{
\begin{array}{ll}  \displaystyle
\left[\sin^{-1}\left(\sqrt{\tau}\right)\right]^2, & (\tau\leq 1) \\
\displaystyle -\frac{1}{4}\left[ \log\left(\frac{1+\sqrt{1-\tau^{-1}}}
{1-\sqrt{1-\tau^{-1}}}\right)-i\pi \right]^2, \hspace{0.5cm} & (\tau>1) \, .
\end{array} \right. 
\label{eq:ftau} 
\end{eqnarray}
%%%%%%%%%%%%%%%%%%%%%%%%%%%%%%%%%%%%%%%%%%%%%%%%%%%%%%%%%%%%%%%%%%%%%%%%%%%%%

\subsection{$H^0_1 \to Z \gamma $}
Similarly the corresponding decay width of the Higgs $(H^0_1)$
into $Z\gamma $ \cite{Cahn:1978nz, Bergstrom:1985hp, Djouadi:1996yq, Chen:2013vi} also occurs at one loop level involving the
exchange of spin 1/2, spin 1 and spin 0 particles in the loops:
\begin{eqnarray}
\Gamma(H^0_1 \to Z \gamma) &=& 
\frac{\alpha M_{H^0_1}^3}{512\pi^4}\left( 1 -\frac{M_Z^2}{M_{H^0_1}^2} 
\right)^3\bigg|
\sum_f\frac{4N_cQ_f}{m_f}g_{Z f \bar f}g_{H^0_1 f \bar f}A^h_{1/2}
(\tau_h^f,\tau_Z^f) \nonumber\\
&+&  \frac{1}{2M_W^2}g_{H^0_1 V V}g_{Z V V}A_1^h(\tau_h^V,\tau_Z^V)
+\frac{N_{c,S}Q_S}{M_S^2}g_{ZSS}g_{H^0_1SS}A^h_0(\tau_h^{S},\tau_Z^{S})
\bigg|^2, 
\end{eqnarray}
where  the Higgs boson couplings to $f,V,S$ are explained before. 
Here, $g_{Z f {\bar f}},\, g_{ZWW} $ and $g_{ZSS}$ represent $Z$ boson couplings
to $f,W^\pm, S$ respectively. We define, $\tau_h^i=4M_i^2/M_{H^0_1}^2, 
\tau_Z^i=4M_i^2/M_Z^2$, where $i$ represent fermion, vector and
scalar particles as defined for $H^0_1 \to \gamma \gamma $ decay process.
The corresponding loop-factors are given by
\begin{eqnarray}
A^h_0(\tau_h,\tau_Z) &=& I_1(\tau_h,\tau_Z),\nonumber \\
A^h_{1/2}(\tau_h,\tau_Z) &=& I_1(\tau_h,\tau_Z)-I_2(\tau_h,\tau_Z), \\
A^h_1(\tau_h,\tau_Z) &=& 4(3-\tan^2\theta_W)I_2(\tau_h,\tau_Z)+ 
%\nonumber \\ &&
\left[(1+2\tau_h^{-1})\tan^2\theta_W-(5+2\tau_h^{-1})\right]I_1(\tau_h,\tau_Z).\nonumber
\end{eqnarray} 
The functions $I_1$ and $I_2$ are given by 
\begin{eqnarray}
I_1(\tau_h,\tau_Z) &=&  \frac{\tau_h \tau_Z}{2\left(\tau_h-\tau_Z\right)} 
+\frac{\tau_h^2\tau_Z^2}{2\left(\tau_h-\tau_Z\right)^2}
\left[f\left(\tau_h^{-1}\right)-f\left(\tau_Z^{-1}\right)\right] \nonumber \\
&& +\frac{\tau_h^2\tau_Z}{\left(\tau_h-\tau_Z\right)^2}
\left[g\left(\tau_h^{-1}\right)-g\left(\tau_Z^{-1}\right)\right],\nonumber\\
I_2(\tau_h,\tau_Z) &=& -\frac{\tau_h\tau_Z}{2(\tau_h-\tau_Z)}\left[f\left(\tau_h^{-1}\right)-f\left(\tau_Z^{-1}\right)\right],
\end{eqnarray}
where the function $f(\tau)$ is defined in Eq.~(\ref{eq:ftau}), and the function $g(\tau)$ is defined as
\begin{eqnarray}
g(\tau)=\left\{
\begin{array}{ll}  \displaystyle
\sqrt{\tau^{-1}-1}\sin^{-1}\left(\sqrt{\tau}\right), & (\tau< 1) \\
\displaystyle 
\frac{1}{2}\sqrt{1-\tau^{-1}}\left[ \log\left(\frac{1+\sqrt{1-\tau^{-1}}}
{1-\sqrt{1-\tau^{-1}}}\right)-i\pi\right], \hspace{0.5cm} & (\tau\geq 1) \, .
\end{array} \right. 
\label{eq:gtau} 
\end{eqnarray}
%%%

\subsection{$H^0_1 \to XX $}
The partial widths of the decay modes which proceed via tree level sub-processes will also 
vary and mostly depend on the choice of the different parameters of the LRSUSY model which govern the 
composition of the $\sim 125$ GeV scalar boson as discussed in  Sec. \ref{sec:model}. This in turn would 
modify its coupling to the fermions and weak gauge bosons and affect its respective partial decay widths.

The superpotential has a term of the form $Y_{u}Q_{L}\Phi_{1}Q_{R}$. Hence the coupling of $\phi_{2}^{0}$ to $b$-quarks is proportional to the top Yukawa coupling. Whenever the SM-like Higgs has a sizable $\phi_{2}^{0}$-component 
the Higgs coupling to $b$-quarks is altered significantly. Depending on the relative sign between the components in the eigenvector, the coupling may either increase significantly or become close to zero (or change sign).

At tree-level the mixing between $\{\phi_1,\chi_2\}$ and $\{\phi_2,\chi_1\}$ is zero \cite{Frank:2011jia}.
Hence the admixture of $\phi_2^0$ in the SM-like Higgs boson comes entirely from loop corrections. The dominant 
contribution comes from third generation quark and squark loops. The contribution of these diagrams is proportional 
to the product of top and bottom Yukawa couplings. This product, and hence the mixing element, is large at large values
of $\tan \beta$. The mixing effect is significant when the second lightest Higgs is non-decoupled, which happens 
at values of $\tan \delta$ close to one.

Since $h\to b\overline{b}$ is the dominant decay mode, a substantial change in this coupling will change 
all other branching ratios. Hence there is an anti-correlation between the $h\to b\overline{b}$ signal strength
and all other signal strengths. The effect is so strong that it leads to a strong correlation between all other 
signal strengths. We highlight this behavior in the next section when we discuss the fit to the Higgs data from our
parameter scans.  The corresponding correction for the Higgs-$\tau$ coupling is proportional to the neutrino Yukawa 
coupling and is hence smaller. Thus the Higgs-$\tau\tau$ coupling is close to the Standard Model prediction.

 Since $\langle\phi^0_2\rangle = 0$, there is no three-point coupling for $\phi_2$ and vector bosons, and when the SM-like
 Higgs has a sizable $\phi_2$-component, the couplings to vector bosons will be suppressed.

%%%%%%%%%%%%%%%%%%%%%%%%%%%%%%%%%%%%
\section{Implications for model parameters}\label{sec:restrictions}
%%%%%%%%%%%%%%%%%%%%%%%%%%%%%%%%%%%%
The  detailed discussion on the left-right supersymmetric spectrum in
Sec. \ref{sec:cspectrum} gives us an idea of the parameters that could 
directly play a significant role in Higgs physics. Therefore we set a number of free parameters to fixed values as shown in Table \ref{fixed:param} and  perform random scans over the ranges of the input parameters shown in Table \ref{input:param}. We divide the choice for fixed 
values and the corresponding scans into three benchmark points while we 
have ensured that the current limits on supersymmetric particles are respected
through all the scans. We also make sure that one of the Higgs mass is always within 
$122~\rm GeV < M_h < 128 ~\rm GeV$ and compare
its properties to the scalar resonance that has been observed at the LHC.

As the LHC experiments have not observed any signal indicative of beyond the SM physics, the most stringently  constrained sector
in almost all extensions of SM is the strongly interacting sector. Therefore we cannot choose very light squark and gluino masses which would be ruled out by experimental data. Note that we have chosen our parameter
space as shown in Tables \ref{fixed:param} and \ref{input:param} where the gluino mass as well as the squark masses are around  $\sim 1$ TeV. A large mixing in the third generation is however still allowed to give a significant splitting for the stop mass eigenstates. In addition, the $SU(2)_R$ breaking scale is also strongly
constrained from direct searches for right-handed gauge bosons ($W_R,Z_R$). We have therefore 
chosen $v_R$ to be sufficiently large to evade the existing mass bounds on such gauge bosons at
the LHC. For the remaining parameters we make sure that our model spectrum is consistent with these experimental constraints \cite{Beringer:1900zz} :
\begin{itemize}
\item  Lower limits on superpartner masses from LEP and the 7 \& 8 TeV run of the LHC. 
\item  Low energy flavour physics processes :
$b \to s \gamma $, $B_s \to \mu^+\mu^-$. 
\end{itemize}
The superpartner masses are controlled by the choices of soft mass parameters and the $\mu$-parameter. We also require the lightest neutralino to be heavier than $m_h/2$ so that invisible Higgs decays are kinematically forbidden.

The flavour constraints are taken care of by requiring $m_A> 300$~GeV and limiting $\tan \beta$ from above. The lightest charged Higgs contributes to $b\to s\gamma$. In type-II 2HDM this leads to a bound $m_{H^{\pm}}> 316$~GeV \cite{Deschamps:2009rh} but in supersymmetric models the contribution from the chargino cancels partially that of the charged Higgs, see \eg \cite{Chen:2009cw, Garisto:1993jc}. The mass of the charged Higgs follows closely the lightest CP-odd Higgs mass as can be seen from Fig. \ref{fig:mH}. The largest contribution to the reaction $B_s \to \mu\mu$ is mediated by the lightest CP-odd Higgs \cite{Babu:1999hn,Isidori:2001fv}. That contribution is at largest at large values of $\tan \beta$. On the other hand, as discussed previously, the Higgs coupling to b-quarks is often altered at large $\tan\beta$ so much that it would be experimentally excluded.

We analyze numerically the various Higgs decay modes in our
model and study to what extent they differ from the SM predictions.
The enhancement or the suppression over the SM can be studied using the signal strengths $(R_{XX})$,  
defined as the Higgs production cross section
times the branching ratio, normalized to the SM value: 
\begin{eqnarray}
R_{XX} =\frac{\sigma(pp \to h)\times BR(h\to XX)}{\sigma(pp \to h)_{SM}\times BR(h\to XX)_{SM}}
\label{signal_rate}
\end{eqnarray}
Here $\sigma(pp \to h)$ gives the on-shell production cross-section of the Higgs boson. 
For the Higgs production modes, we have considered gluon-gluon fusion ($gg$) and the
associated vector boson production ($VH$), where $V$ stands for $W$ or $Z$ boson.
While for most of the decay channels, $gg$ production mode is considered, we use the $VH$ mode for
$b\bar b$ and $\tau^+\tau^-$ final states. 

The Higgs production cross-section through $gg$ fusion is calculated both in the 
LRSUSY and in the SM, using the publicly available package {\tt HIGLU} \cite{Spira:1996if}, at LHC with the 
center of mass energy, $\sqrt{s} = 8$ TeV. 
As the cross section for the Higgs production implemented in {\tt HIGLU} is for the SM, we 
calculated the effective coupling strengths in the LRSUSY model and obtained the cross sections by 
modifying the strengths in the {\tt HIGLU} code.  The partial decay
widths evaluation for the Higgs is done using another publicly available package {\tt HDECAY} \cite{Djouadi:1997yw}.
A similar technique to what was done in {\tt HIGLU} is used here too to get the partial widths in the 
LRSUSY model. However, the loop induced decay modes, $h \to \gamma\gamma$
and $h \to Z\gamma$ involve new particles modifying the loop amplitudes, and here we 
used our Fortran code. Note that {\tt HIGLU} includes the QCD correction while calculating the Higgs production
cross-section. Any explicit implementation of QCD or EW corrections are not taken into account 
in calculating the partial decay widths of
$H_1^0 \rightarrow \gamma \gamma$ or $H_1^0 \rightarrow Z \gamma$ modes. However, we expect that the EW contribution to these decays will be the same as in the SM and thus cancel out when we consider the signal strengths. 

\begin{table}[ht!]
\begin{ruledtabular}
\begin{tabular}{c c c c}
Higgs decay channel & Experiment & $\sqrt{s}$ (TeV) (Luminosity in $fb^{-1}$)& Signal strengths \\
\hline
$h \to \tau \bar \tau$ & ATLAS & 8(20.3) & $1.4^{+0.5}_{-0.4}$ \\
$h \to \tau \bar \tau$ & CMS   & 7(5.1) + 8(19.7) & $0.91^{+0.27}_{-0.27}$ \\ \hline
$h \to b\bar b$ & ATLAS(VH) & 7(4.7) + 8(20.3) & $0.2^{+0.7}_{-0.6}$ \\
$h \to b\bar b$ & CMS(VH) & 7(5.1) + 8(18.9) & $0.93^{+0.49}_{-0.49}$ \\ \hline
$h \to WW^*$ & ATLAS & 7(4.6) + 8(20.7) & $1.00^{+0.32}_{-0.29}$ \\
$h \to WW^*$ & CMS & 7(5.1) + 8(19.7) & $0.83^{+0.21}_{-0.21}$ \\ \hline
$h \to ZZ^*$ & ATLAS & 7(4.6) + 8(20.7) & $1.44^{+0.40}_{-0.35}$ \\
$h \to ZZ^*$ & CMS & 7(5.1) + 8(19.7) & $1.00^{+0.29}_{-0.29}$ \\ \hline
$h \to \gamma\gamma$ & ATLAS & 7(4.8) + 8(20.7) & $1.57^{+0.33}_{-0.29}$ \\
$h \to \gamma\gamma$ & CMS & 7(5.1) + 8(19.6) & $1.13^{+0.24}_{-0.24}$ \\
\end{tabular}
\end{ruledtabular}
\caption{The measured values for the Higgs signal strength, and their corresponding
uncertainties (lower/upper edges of 1$\sigma$ error bars), in the various channels \cite{ATLAS:2014, Khachatryan:2014ira, CMS:2014}.}
\label{expdata}
\end{table}
We list the observed strengths for the Higgs signal in different final state modes by 
the ATLAS \cite{ATLAS:2014} and CMS \cite{Khachatryan:2014ira,CMS:2014} experiments in Table~\ref{expdata}. 
It is worth mentioning here that we found that there exists enough model (LRSUSY) parameter space 
within the $2\sigma$ limit of the ATLAS and CMS results of Higgs signal strengths. To illustrate our findings, we can choose either the CMS 
or the ATLAS results. The generic features of the fit would 
remain very similar and a definite shift is found in the allowed parameter 
space when using the ATLAS data as the central values and the associated 
errors for the different channels in the two experiments do allow for a wide 
range of signal strength values. However there is a definite overlapping 
region of the parameter space which is common and satisfies data from both 
experiments. As an extension of our findings and for better understanding of 
the model, we first study the variations of few of the relevant 
sparticle masses with $\rm tan \beta$, for which the signal strengths 
$R_{WW^*}$ and $R_{\gamma \gamma}$ are simultaneously within the 
$2\sigma$ error bar of CMS result (see Table.~\ref{expdata}). We do this 
for all the three benchmark points to study the parameters that would 
affect the signal strengths in a significant way. 
\begin{figure}[ht!]
\centering
\begin{subfigure}[b]{0.5\textwidth}
%\begin{tabular}{cc}
\includegraphics[trim = 50 140 40 60,clip,width=3in,height=3.5in]{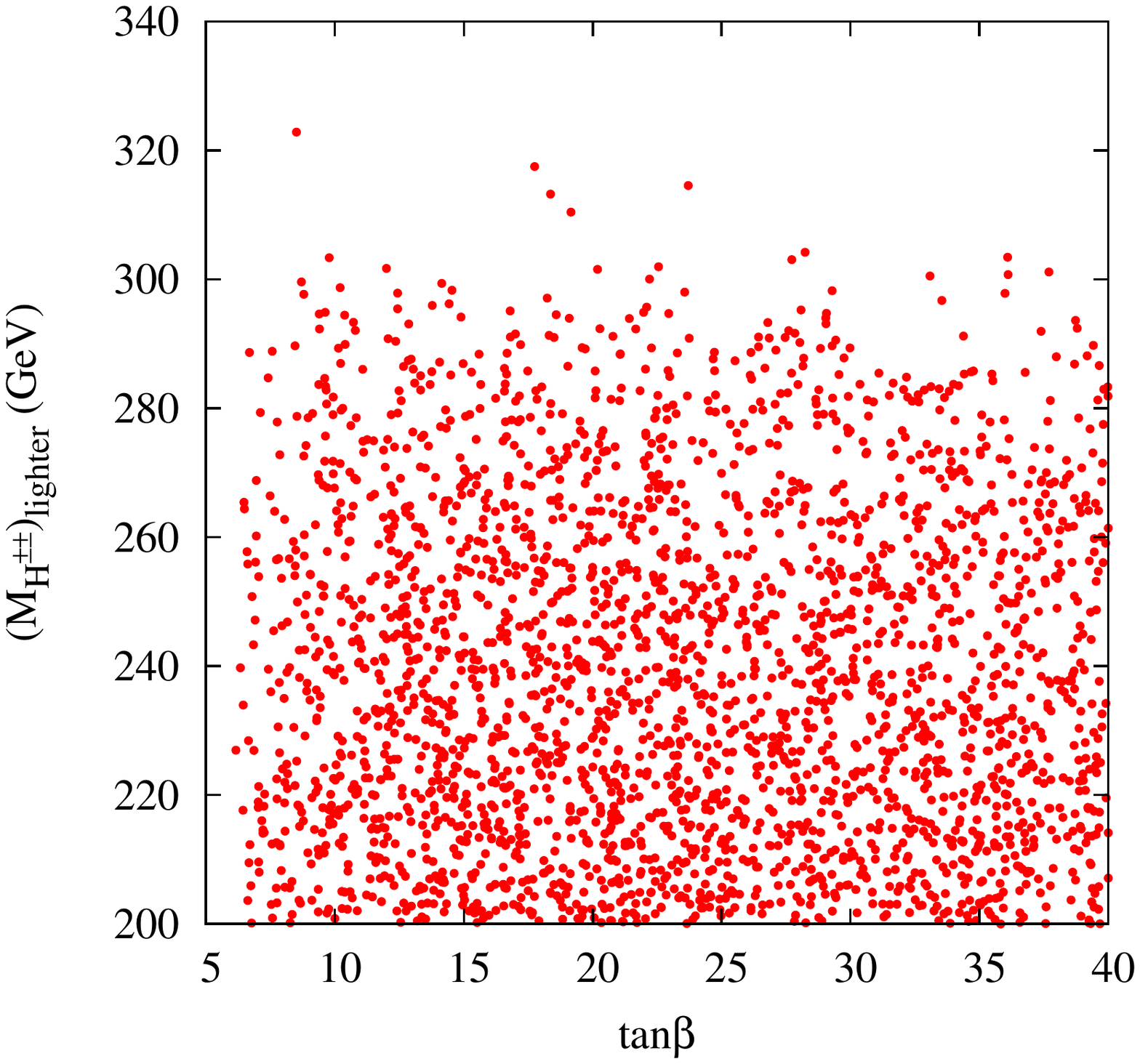} 
\vspace*{-1.2cm}
\caption{}
\end{subfigure}~
\begin{subfigure}[b]{0.5\textwidth}
 %& \hspace*{-0.5cm}
\includegraphics[trim = 50 140 40 60,clip,width=3in,height=3.5in]{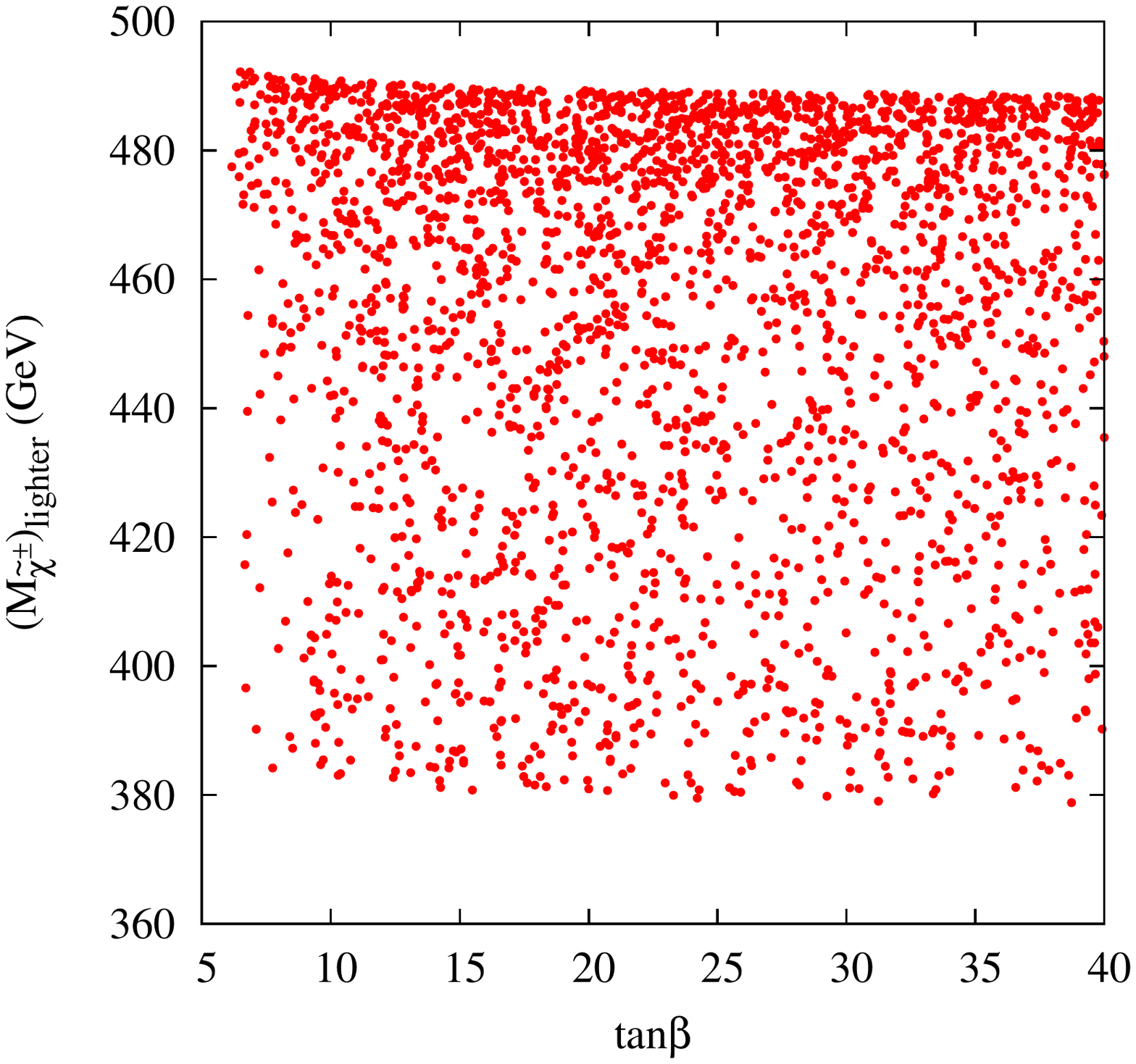}
\vspace*{-1.2cm}
\caption{}
\end{subfigure}~
\vskip -0.1in
\caption{Allowed parameter space for benchmark BP3 in the (a) ${\rm tan}\beta-{\rm M_{H^{\pm \pm}}}$ and (b) ${\rm tan}\beta-{\rm M_{\tilde \chi^{\pm}}}$ plane, 
where $\rm R_{\gamma\gamma}$ and $\rm R_{WW^*}$ are within $2\sigma$ of CMS result.}
\label{tb:mldch}
\end{figure}

We choose the $R_{WW^*}$ and $R_{\gamma \gamma}$ results, since these provide
the most stringent bound on the parameter space. 
In Fig.~\ref{tb:mldch}(a), we show the allowed values for the mass of the
doubly charged Higgs mass ($M_{H^{\pm\pm}}$) as a function of $\rm tan \beta$ for the BP3 scenario. Note that the variation in $\tan\beta$ 
does not play any significant role for the doubly charged scalar mass, but 
does affect the signal strength for the 125 GeV Higgs in the fermionic decay modes. This in turn would affect the branching in the $\gamma\gamma$ mode where the light doubly charged scalar contributes in the loop. The Higgs mass and the resulting bound on $\tan \beta$ have an ef\mbox{}fect on the coupling between the SM-like Higgs and the doubly charged Higgs bosons, which can be seen as follows. If we write the lighter doubly charged Higgs state as $H_1^{++} \equiv a~\delta^{c++}+b~\bar{\delta}^{c++}$, the coupling between the SM-like Higgs and the doubly charged Higgs is
\begin{equation}
g_{hH^{++}H^{--}}=-\sqrt{2}ab\frac{\lambda v \mu_{\mathrm{eff}}}{v_s}\sin 2\beta +\frac{1}{4}(a^2-b^2)g_R^2 v.
\end{equation}
The lighter state is almost the symmetric combination with values around $a=0.73$ and $b=0.69$. Hence the term $a^2-b^2$ is quite small. At large or even moderate values of $\tan \beta$,  the value of $\sin 2\beta$ is small so the coupling will be quite limited. This leads to an interesting observation in the LRSUSY model.
We find here that the doubly charged scalar contribution is therefore comparatively less than that of the
singly charged scalars in both $H \to \gamma \gamma $ and $H \to Z \gamma $
processes, simply because the $h-H^{\pm\pm}-H^{\mp\mp}$ coupling is weaker
than $h-H^{\pm}-H^{\mp}$ coupling. A relative suppression at the level of relative coupling strength
is found to be $g_{h H^{\pm\pm}H^{\mp\mp}}/g_{h H^+H^-}\approx 1/20 $ and this makes 
the doubly charged contribution substantially smaller than singly charged one. This behavior is 
completely opposite to  Type-II Seesaw models \cite{Arhrib:2011vc, Chun:2012jw, Dev:2013ff}, 
where the largest contribution to these one-loop processes comes 
from the virtual exchange of doubly charged scalars. 

Due to the form of the tree-level bound it is very dif\mbox{}f\mbox{}icult to have a Higgs mass around $125$~GeV when $\tan \beta$ is close to 1, where the coupling between the Higgs and doubly charged Higgs bosons would be large.
Therefore an 
indirect dependence on its mass can be obtained as a function of $\tan\beta$ 
here. As discussed in Sec. \ref{sec:cspectrum}, a large value of the soft 
parameter $(M_{\tilde \ell})_{R}$ helps in raising the doubly charged Higgs mass through radiative corrections, to above its
current experimental limits of 200 GeV, provided it decays with 100\% probability into $\tau^\pm\tau^\pm$.
We also note that the upper limit of the range over which this parameter is scanned for 
benchmark BP3 is much lower than that for the other two benchmark points. Thus we find that 
much lighter doubly charged scalars are preferred and also satisfy the CMS data. Very similar
feature is observed for BP1 and BP2 cases, but as the scan range over $(M_{\tilde \ell})_{R}$ is for larger values, we also get heavier doubly charged states in the spectrum for these benchmark points. In Fig.~\ref{tb:mldch}(b) we show the lightest chargino mass as a function of 
$\tan\beta$. In this case, there is a dependence on $\tan\beta$ as seen 
from Eq. \ref{eq:mchargino}. However the dominant parameter is the value for parameters $M_2$ and $\mu_{eff}$ which set the upper limit of the lighter chargino mass to $\sim$200 GeV for BP2 and $\sim$500 GeV for BP1 and BP3 scenarios respectively, and we observe that the bounds from Higgs signal strengths allow almost all the available mass region. This is illustrated for BP3 case in  Fig.~\ref{tb:mldch} (b). 
%%%
\begin{figure}[ht!]
\centering
\begin{subfigure}[b]{0.5\textwidth}
%\begin{tabular}{cc}
\includegraphics[trim = 50 140 40 60,clip,width=3in,height=3.5in]{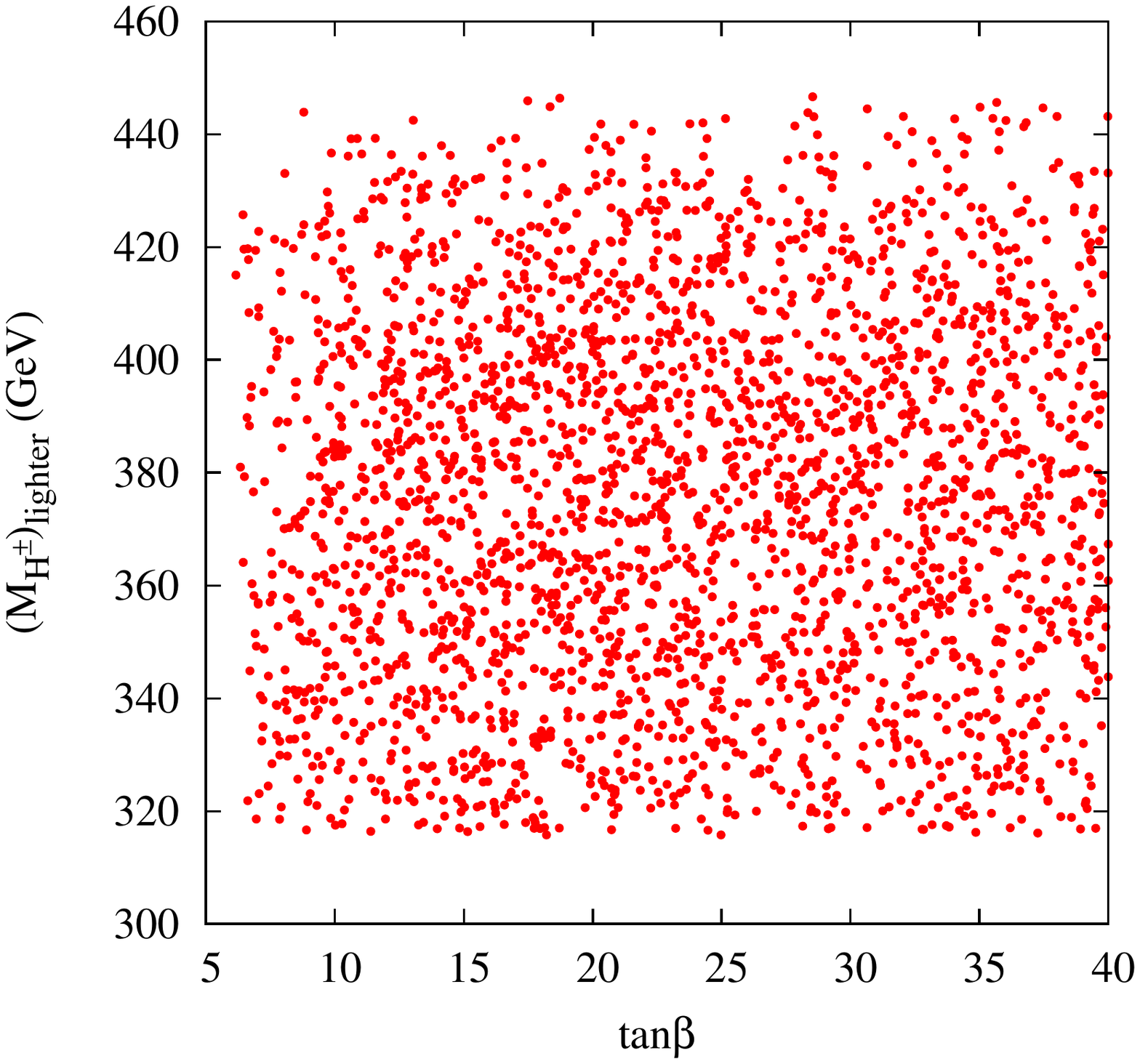} 
\vspace*{-1cm}
\caption{}
\end{subfigure}~
\begin{subfigure}[b]{0.5\textwidth}
 %& \hspace*{-0.5cm}
\includegraphics[trim = 50 140 40 60,clip,width=3in,height=3.5in]{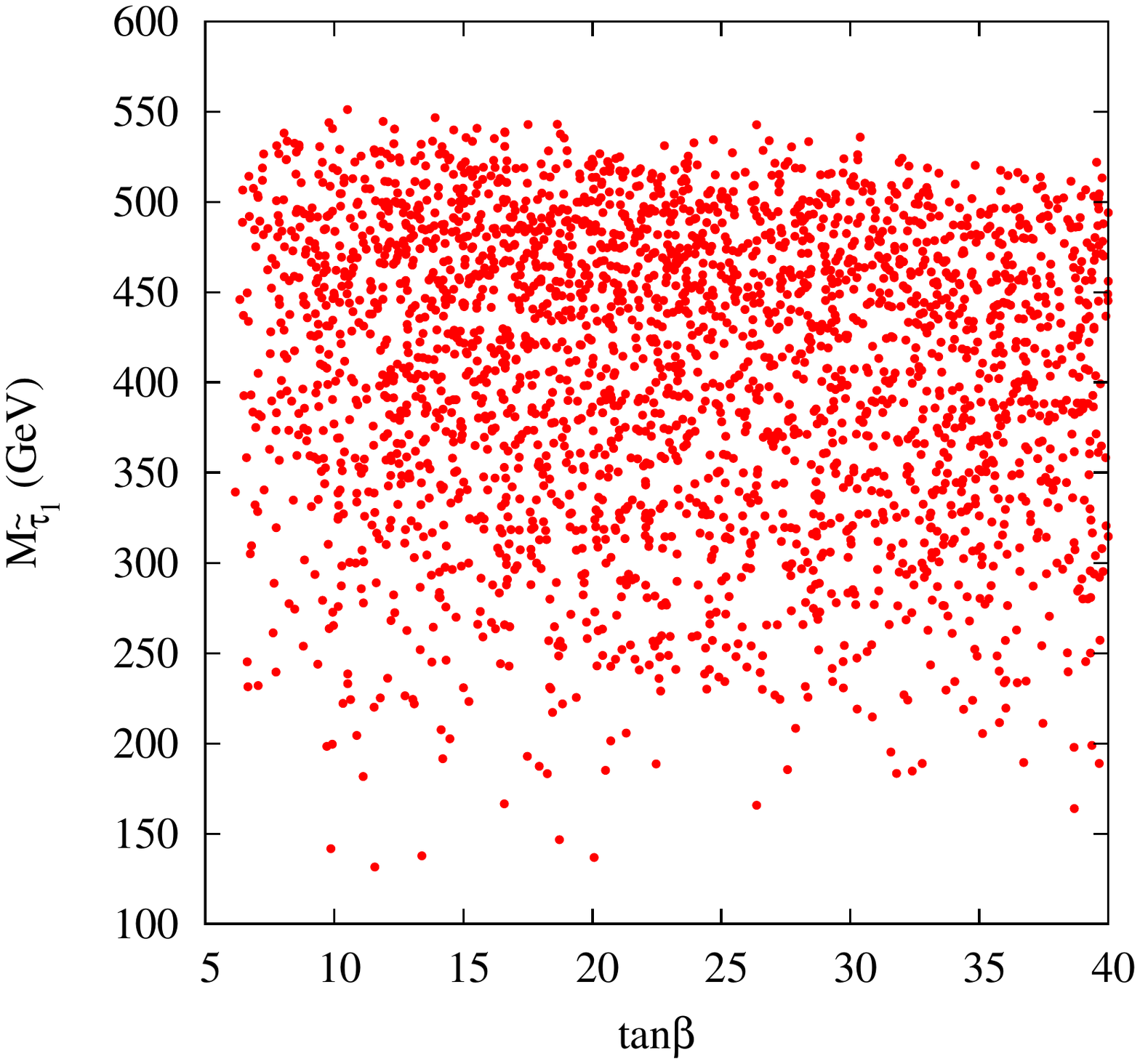}
\vspace*{-1cm}
\caption{}
\end{subfigure}~
\vskip -0.1in
\caption{Allowed parameter space for BP3 scenario in the (a) ${\rm tan}\beta-{\rm M_{H^\pm}}$ and (b) ${\rm tan}\beta-{\rm M_{\tilde \tau_1^{\pm}}}$ plane, where $\rm R_{\gamma\gamma}$ and $\rm R_{WW^*}$ are within $2\sigma$ of CMS result.}
\label{tb:mlch}
\end{figure}
In Fig.~\ref{tb:mlch} we show the variation of lighter singly charged
Higgs boson mass with $\rm tan \beta$ as well as the lightest stau ($\tilde{\tau}_{1}$) mass for the benchmark point BP3. It is important to 
note that, while for BP1 and BP2 cases, the lighter singly charged Higgs mass can vary from 300 GeV to 650 GeV, it is in the range 320-450 GeV for BP3 scenario. As discussed before, and also seen in Fig.~\ref{fig:mH}, a large $\tan\delta$ value yields a lighter singly charged scalar in the spectrum. This is the case in BP3 where the scan runs over larger values of $\tan\delta$. We checked 
for the consistency of such light charged scalars with flavour physics constraints and we find that the Higgs data gives a much weaker constraint compared to flavour physics limits, which we include.  
%{\color{red} I believe this statement is correct, and otherwise much lighter single charged Higgs would be allowed by the Higgs data}. 
We also find that a much 
lighter $\tilde{\tau}_1$ is allowed, consistent with our choice of parameters for benchmark BP3, and the Higgs data does not constrain them very much either. 
Note that $\tilde{\tau}_1$ is significantly heavier for the other benchmark BP1 and BP2 due to the parameter choice for the scan of the slepton soft mass parameters. Thus we find that a large region of the parameter space in the left-right supersymmetric scenario still survives  
\begin{figure}[ht!]
\centering
\begin{subfigure}[b]{0.5\textwidth}
\caption{}
\vspace*{-1.5cm}
%\begin{tabular}{cc}
\includegraphics[trim = 50 80 40 80,clip,width=3in,height=3.5in]{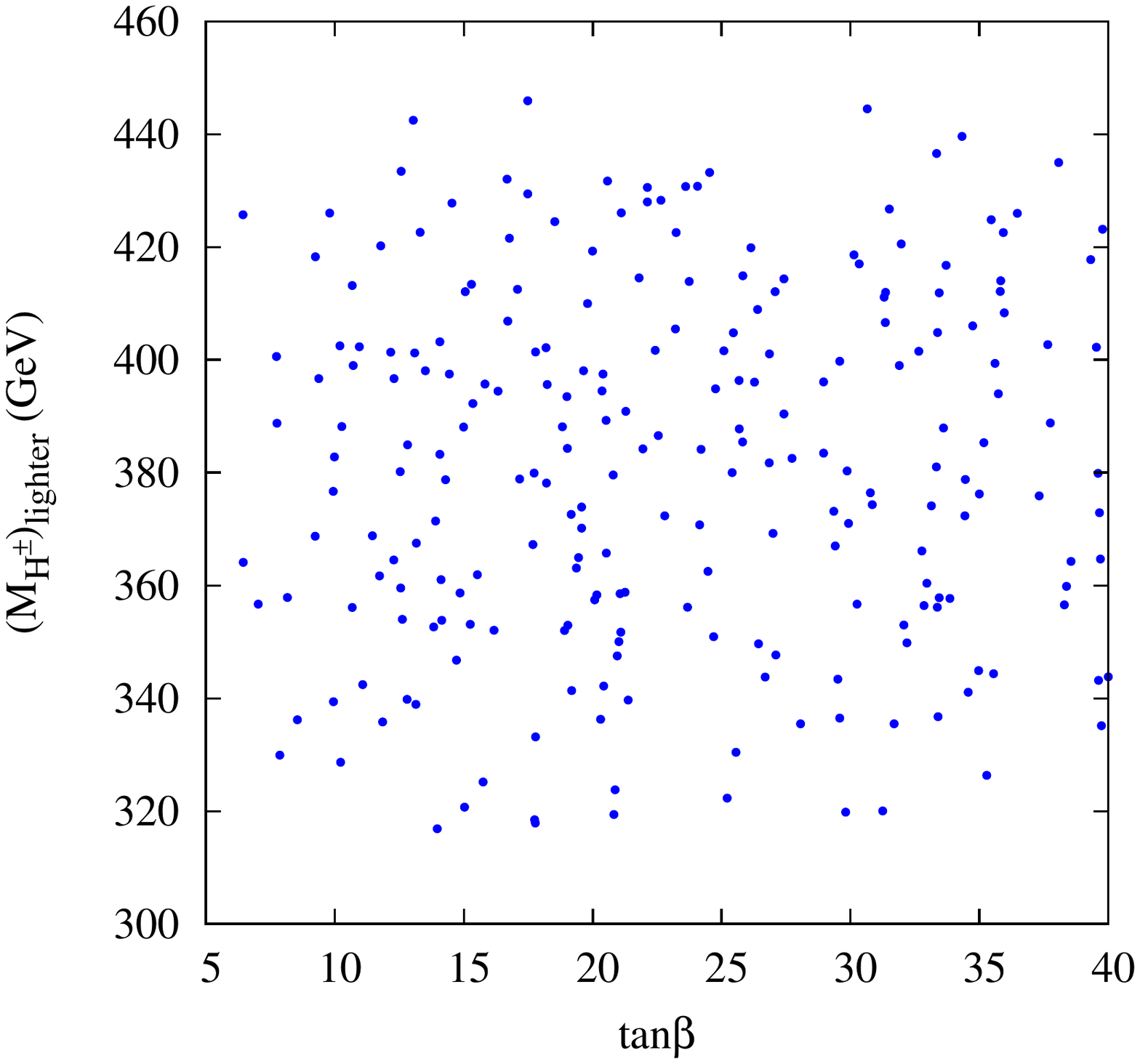}
\vspace*{-2.5cm}
\end{subfigure}~
\begin{subfigure}[b]{0.5\textwidth}
 %& \hspace*{-0.5cm}
\caption{}
\vspace*{-1.5cm}
\includegraphics[trim = 50 80 40 80,clip,width=3in,height=3.5in]{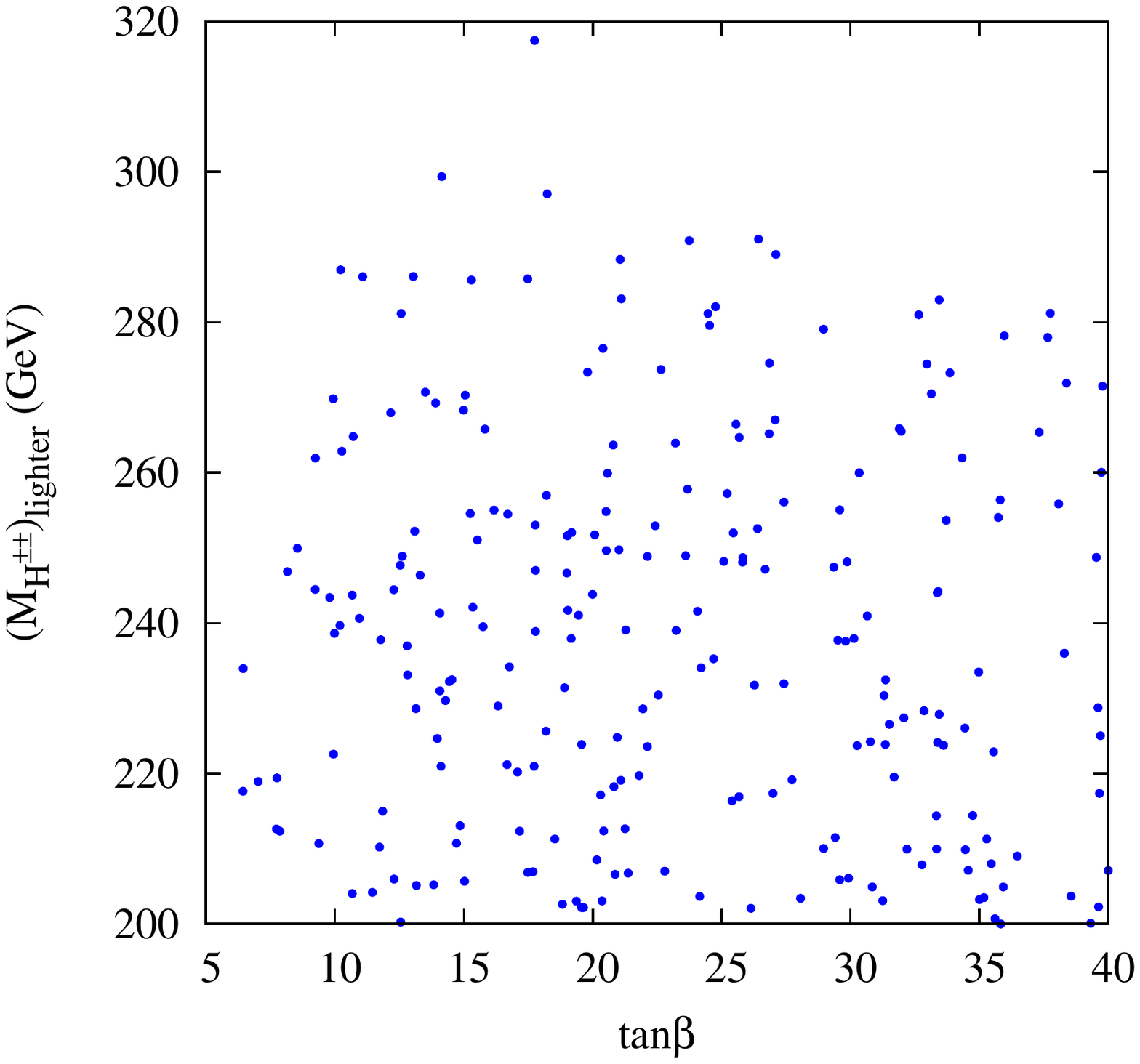} %\\
\vspace*{-2.5cm}
%\caption{}
\end{subfigure}~ \\
\vspace*{1.0cm}
\begin{subfigure}[b]{0.5\textwidth}
\caption{}
\vspace*{-1.5cm}
\includegraphics[trim = 50 80 40 80,clip,width=3in,height=3.5in]{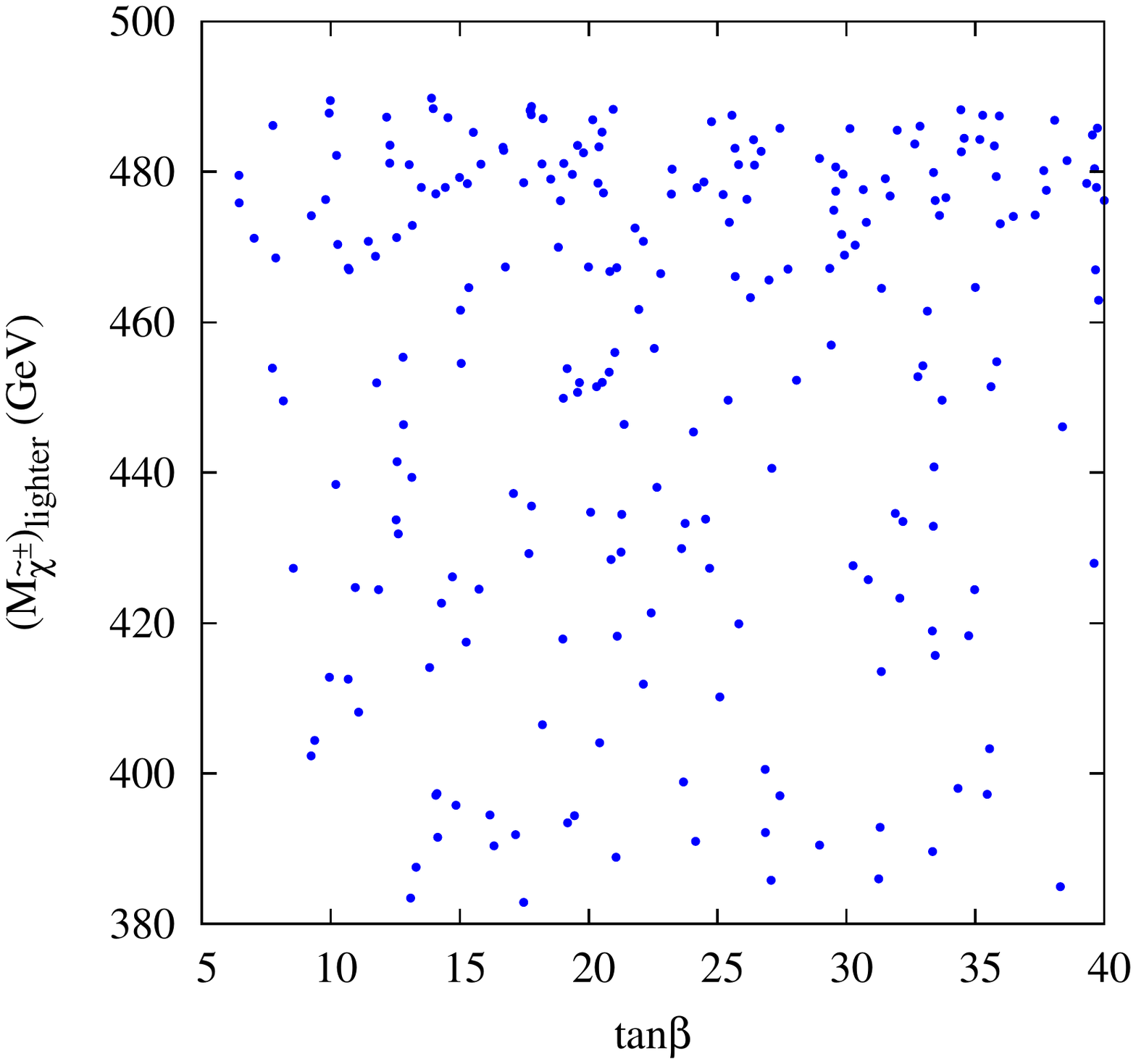}~~ %& \hspace*{-0.5cm}
\end{subfigure}~
\begin{subfigure}[b]{0.5\textwidth}
\caption{}
\vspace*{-1.5cm}
\includegraphics[trim = 50 80 40 80,clip,width=3in,height=3.5in]{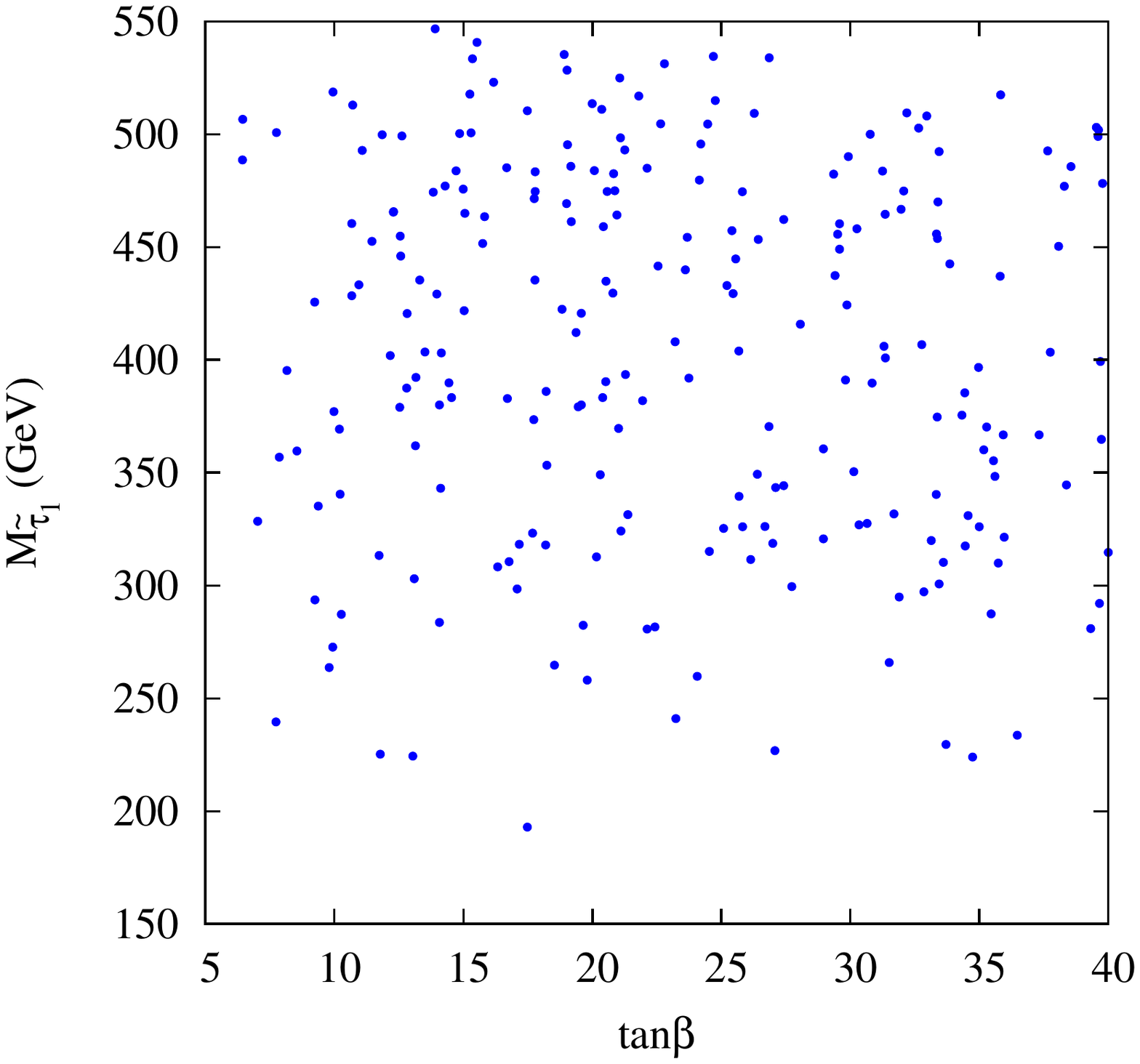} 
\end{subfigure}~
%\end{tabular}
\vskip -0.5in
\caption{Allowed parameter space for benchmark point BP3 in the plane ${\rm tan}\beta$ and (a)${\rm M_{H^\pm}}$
(b)${\rm M_{H^{\pm\pm}}}$, (c) ${\rm M_{\tilde \chi^{\pm}}}$ , (d) ${\rm tan}\beta-{\rm M_{\tilde \tau^{\pm}}}$,
where $\rm R_{\gamma\gamma}$ and $\rm R_{WW^*}$
are within $2\sigma$ of CMS result and the points indicate those values for which the lightest CP-even Higgs coupling is within $2\%$ of SM value.}
\label{compare:parameter}
\end{figure}
%%%
when confronted by the Higgs data at LHC. Light singly- and doubly-charged scalars,
sleptons (benchmark BP3) and charginos (benchmark BP2) are all viable and agree with the Higgs signal strengths. Of course when one considers the parameter scan, it is also imperative to view the scan where the SM coupling strengths are not altered 
by large values. There can be two ways of achieving this. The most likely 
case would be the case where the SM sector is completely decoupled and none 
of the supersymmetric particles contribute to the Higgs decay. The other and more interesting option would be to consider the non-decoupling scenario, where  the supersymmetric particles conspire in their contributions to give
similar coupling strengths as the SM Higgs. This would mean that a light left-right supersymmetric spectrum would still coexist and is waiting to be discovered at the LHC. 

It is therefore interesting to check the presence 
of the LRSUSY model even if the Higgs coupling is almost SM like. Therefore, for illustration purposes, we set the lightest CP-even Higgs couplings to all SM particles within $2\%$ of the SM value and calculate the 
Higgs signal strengths\footnote{We should however point out that such a high precision for the measurement of all the Higgs couplings would be extremely difficult at the LHC even at very high luminosity.}. We analyze this by studying the allowed parameter space only for benchmark point BP3.  A few things worth noting here is that although the $ht\bar{t}$ and $hWW$ couplings are not affected at all throughout the scan, the $hb\bar{b}$ is affected significantly, as discussed
earlier. So we find this to be the dominating factor in allowing only a 2\% shift. As can be seen from Fig.~\ref{compare:parameter}, there still exists 
a strong possibility of having a light LRSUSY spectrum,  even if the Higgs couplings are very close to the SM values. The blue points in Fig.~\ref{compare:parameter} show that quite a significant parameter range  of LRSUSY which is light and can modify 
the Higgs decays to conspire and give SM like strength for the $\sim 125$ GeV scalar state for the observed decay modes. Or it is equally possible to have these light states which do not shift the Higgs couplings by large amounts but will show up in other complementary channels through direct 
production at the LHC running with high enough luminosity.

%%%%%%%%%%%
\begin{figure}[h!]
\centering
$\begin{array}{ccc}
\includegraphics[trim = 50 120 40 60, width=3in,height=3.5in]{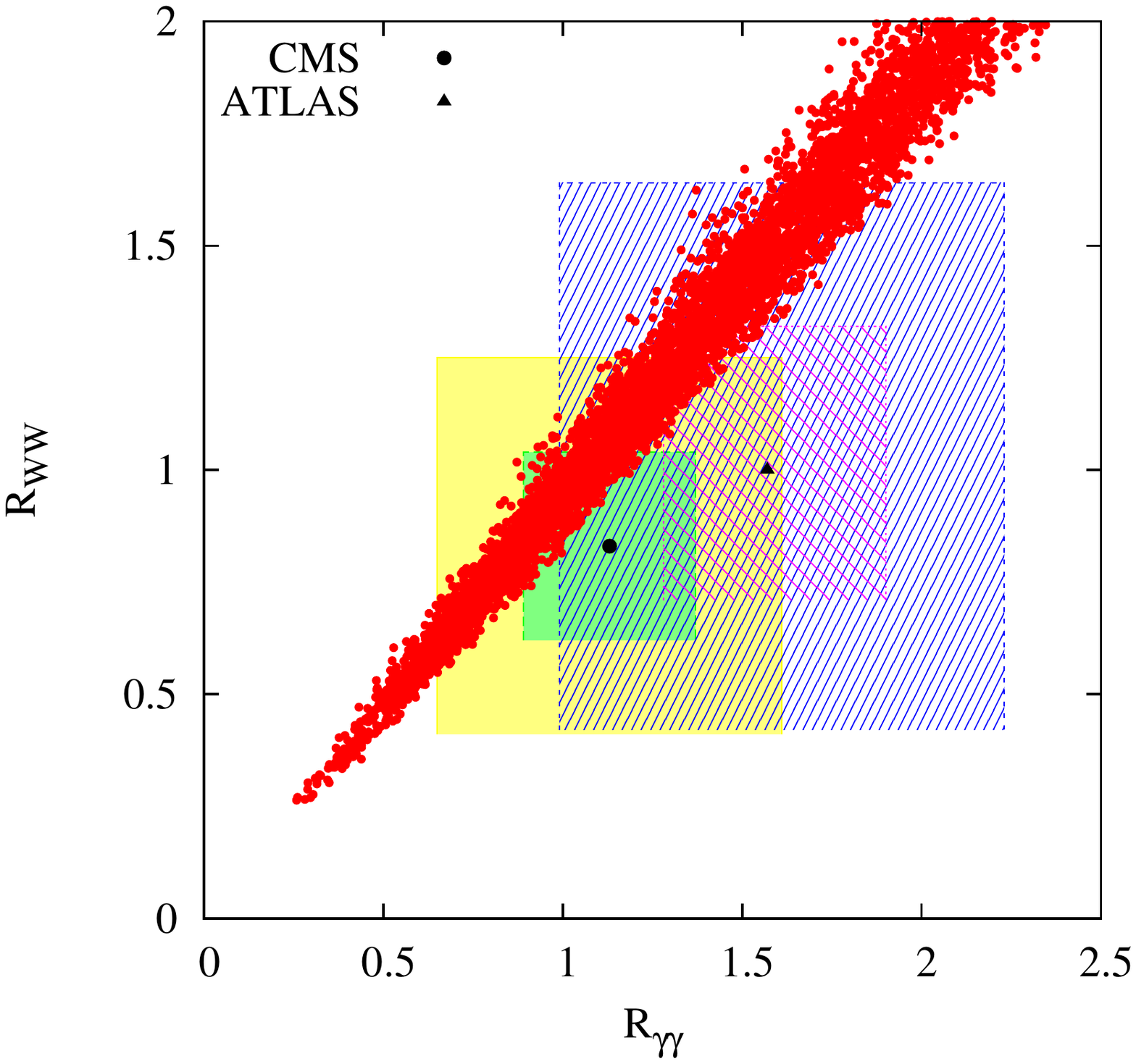}
\includegraphics[trim = 50 120 40 60, width=3in,height=3.5in]{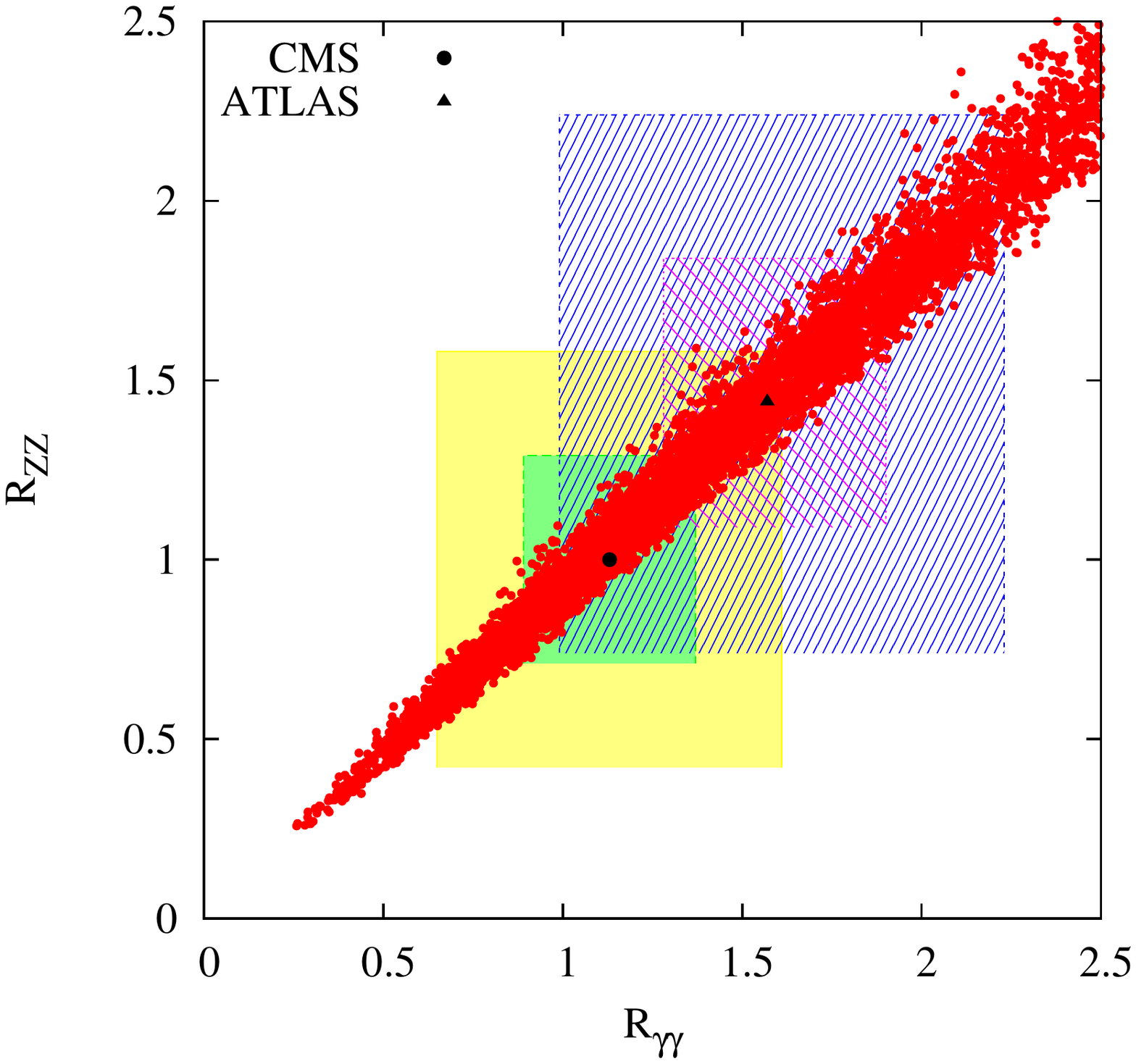}
\end{array}
$
\vskip-0.3in
\caption{Correlation between signal strengths for the Higgs boson in the $\gamma\gamma-WW^*$ (left panel)
 and $\gamma\gamma-ZZ^*$(right panel) channel in the LRSUSY model for the benchmark BP3. 
The corresponding central values as observed by the CMS (circular black point) and ATLAS (triangular black point) Collaborations in the two channels are shown. 
We also include the associated $1\sigma$ and $2\sigma$ deviations allowed by the data, shown as overlying rectangles around the central values. For details see the text.}
\label{fig:Rggww}
\end{figure}
\begin{figure}[h!]
\centering
\vskip-0.3in
$\begin{array}{ccc}
\includegraphics[trim = 50 120 40 60, width=3in,height=3.5in]{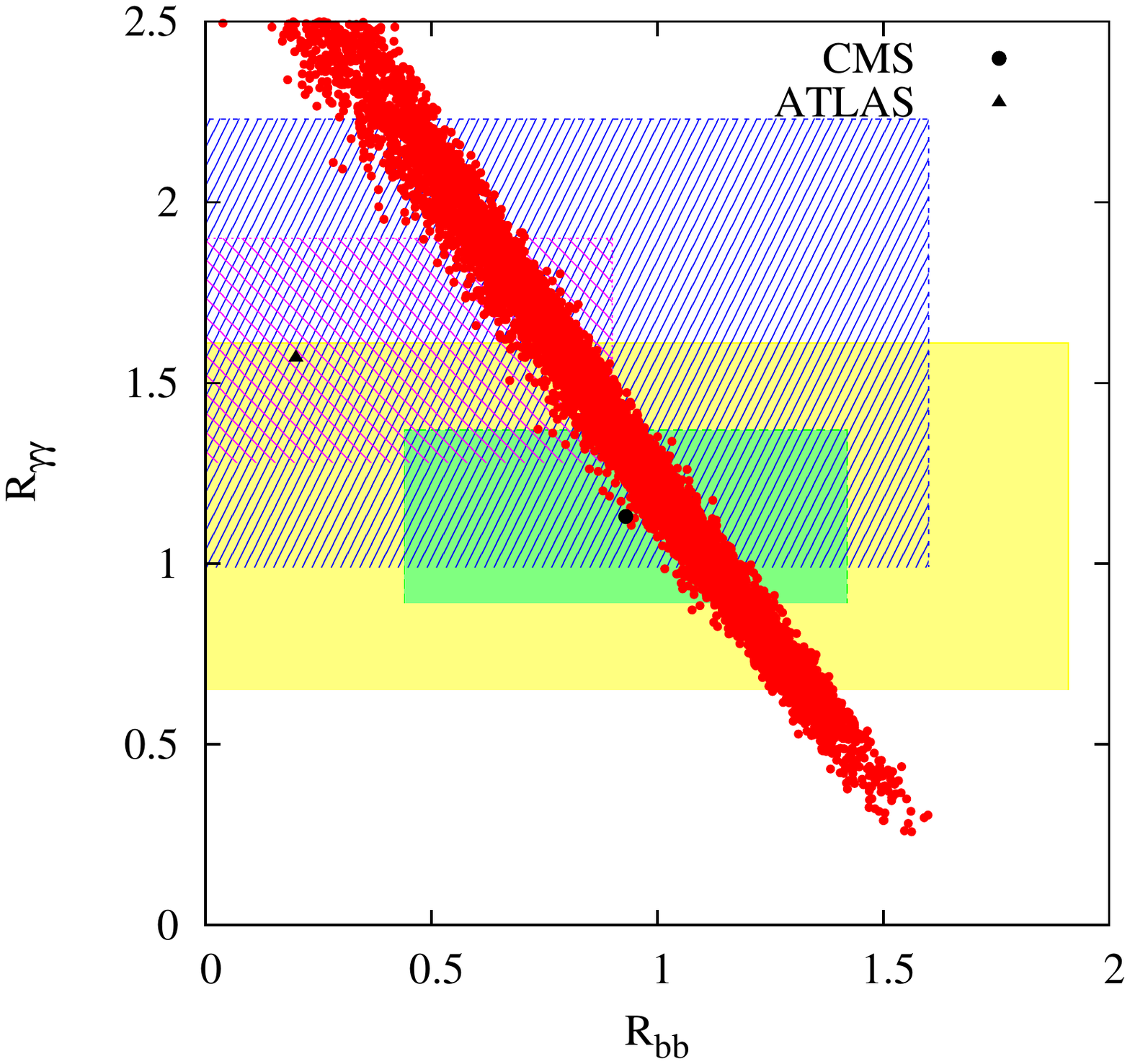}
\includegraphics[trim = 50 120 40 60, width=3in,height=3.5in]{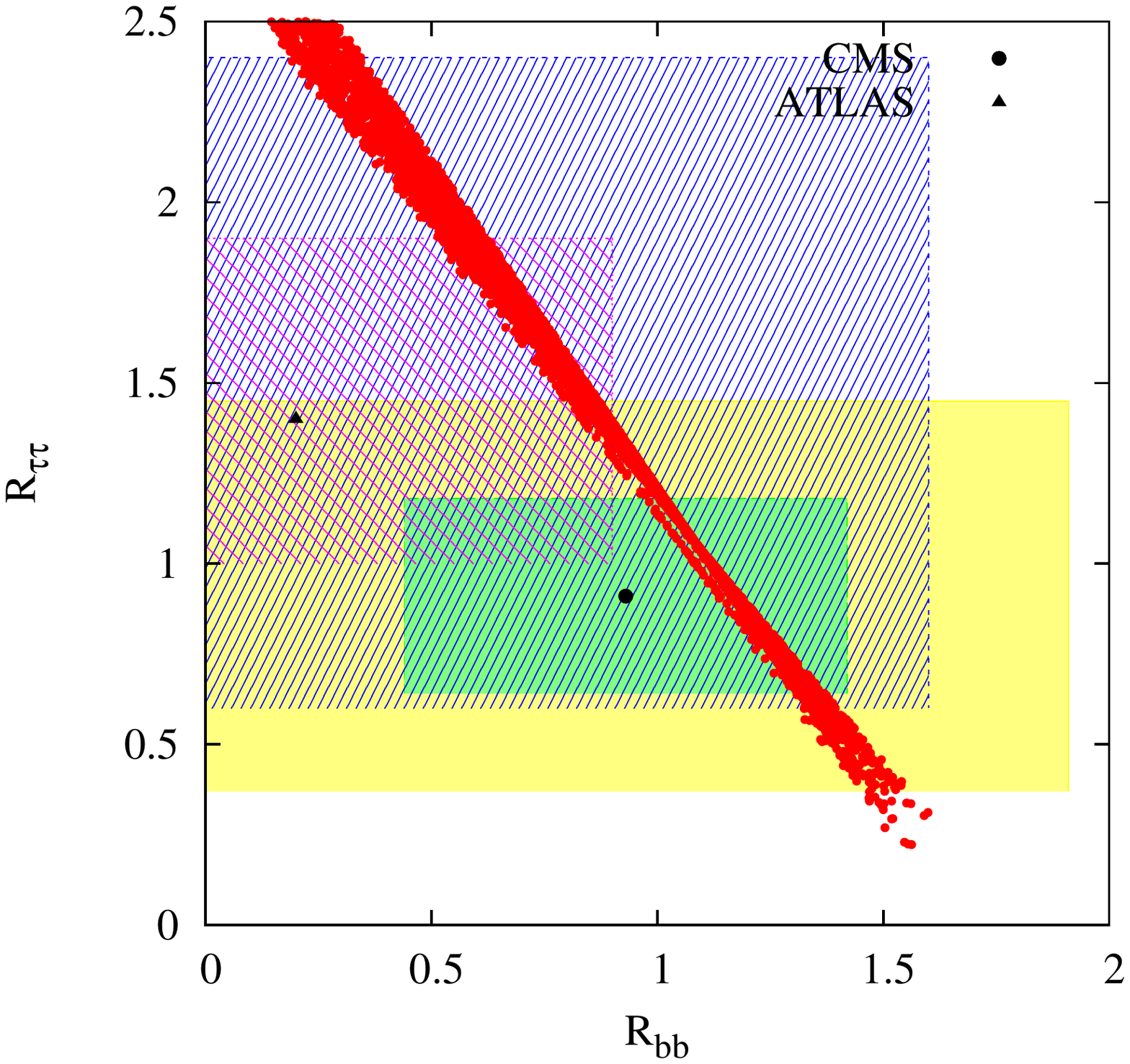}
\end{array}
$
\vskip-0.3in
\caption{Correlation between signal strengths for the Higgs boson in the
$b \bar b$, $\gamma\gamma$ and $\tau \bar \tau$
channel in the LRSUSY model for benchmark BP3. Color
specifications are same as Fig.~[\ref{fig:Rggww}].}
\label{fig:Rggbb}
\end{figure}

%%%%%%%%%%%%%%
We now focus on the signal strengths that we obtain through our scan over the 
different parameters in the LRSUSY model. Figs. \ref{fig:Rggww} and 
\ref{fig:Rggbb} show the correlations among the various signal strengths. The 
scanned parameter points in the LRSUSY model are shown in red.
The black 
solid triangle represents the best-fit point of the ATLAS Collaboration, with the
patterned blue and magenta patches showing regions of $1\sigma$ and $2\sigma$ 
uncertainty respectively, while the black 
circle denotes the best-fit value of the signal rates predicted by the 
CMS Collaboration with the solid green and yellow patches show the 
$1\sigma$ and $2\sigma$ uncertainty around it, respectively.
In Table. \ref{expdata}, we give the 
experimental values of the signal rates for different Higgs decay channels 
with the corresponding center of mass energy and integrated luminosity. 
When compared to the mass plots which only use the CMS data, we do find that 
there exists a much extended parameter space for the LRSUSY model which is
allowed by current Higgs data. The ATLAS results in fact give a much larger 
acceptance for the parameter space when compared to the CMS. We find regions 
where the couplings are consistent with the SM values as well as regions where they can significantly vary and modify the signal strengths beyond the SM expectations. Note that both $R_{WW}$ and $R_{ZZ}$ are almost linearly 
correlated to $R_{\gamma\gamma}$. However that is not the case when one considers the $R_{bb}$ shown in Fig. \ref{fig:Rggbb}.
%%%%%%%%%%%%%%%%%%%%%%%%%%%
We find that $R_{bb}$ is anti-correlated to all 
the other signal rates, as also evident from Fig.~\ref{fig:Rggbb}. This is simply because the Higgs boson decays mostly into $b\bar b$ final 
states and hence the total decay width is dependent sensitively on the partial decay width 
$\Gamma(H \to b\bar b)$. Hence, an increase in $b \bar b$ branching ratio will effectively reduce the 
branching ratios of sub-dominant decay channels. All the other rates show strong correlations
among each other. Hence, we see that our model provides a large parameter space greatly consistent with the present LHC data.  

%
%%%%%%%%%%%
\begin{figure}[h!]
\centering
\includegraphics[trim = 50 120 40 60, width=3.4in,height=3.6in]{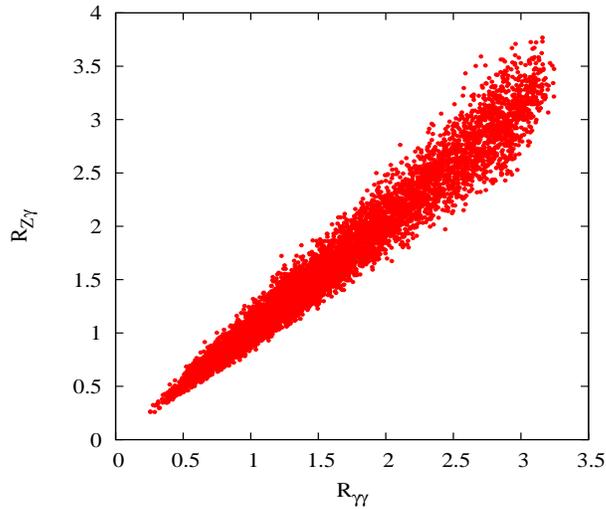}
\vskip-0.3in
\caption{Predicted correlation between signal strengths for the Higgs in the $\gamma\gamma$
and $Z\gamma$ channel in the LRSUSY model for the benchmark BP3 at the LHC with center of mass energy of 14 TeV.}
\label{fig:Rggzga}
\end{figure}
%%%%%%%%%%%%%%
With the LHC expected to run with greater energy and gather data with higher integrated luminosity, it 
is quite clear that it will also be able to measure the Higgs signal in other channels which it could not at the
$\sqrt{s}=7$ and 8 TeV. One such mode would be the remaining loop mediated decay channel $Z\gamma$.
As the LRSUSY model also affects that mode, similar to the $\gamma\gamma$, we present the expected 
correlation between the signal strengths of the Higgs in the two modes $R_{\gamma\gamma}$ and 
$R_{Z\gamma}$ at the 14 TeV run of LHC in Fig. \ref{fig:Rggzga} for the benchmark point BP3.

\begin{figure}[h!]
\centering
\vskip-0.3in
$\begin{array}{ccc}
\includegraphics[trim = 50 140 40 60, width=3.0in,height=3.8in]{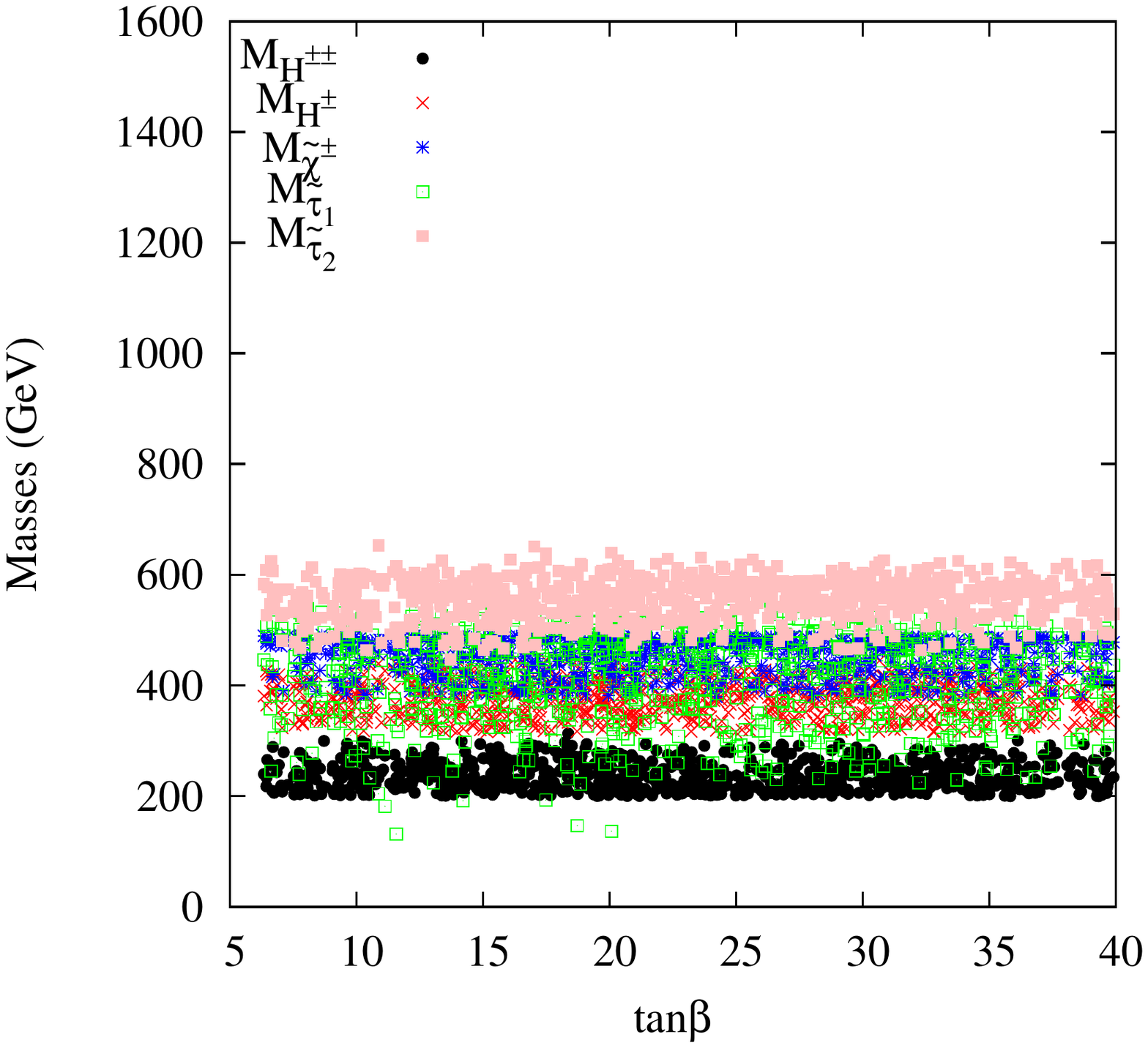}
\includegraphics[trim = 50 140 40 60, width=3.0in,height=3.8in]{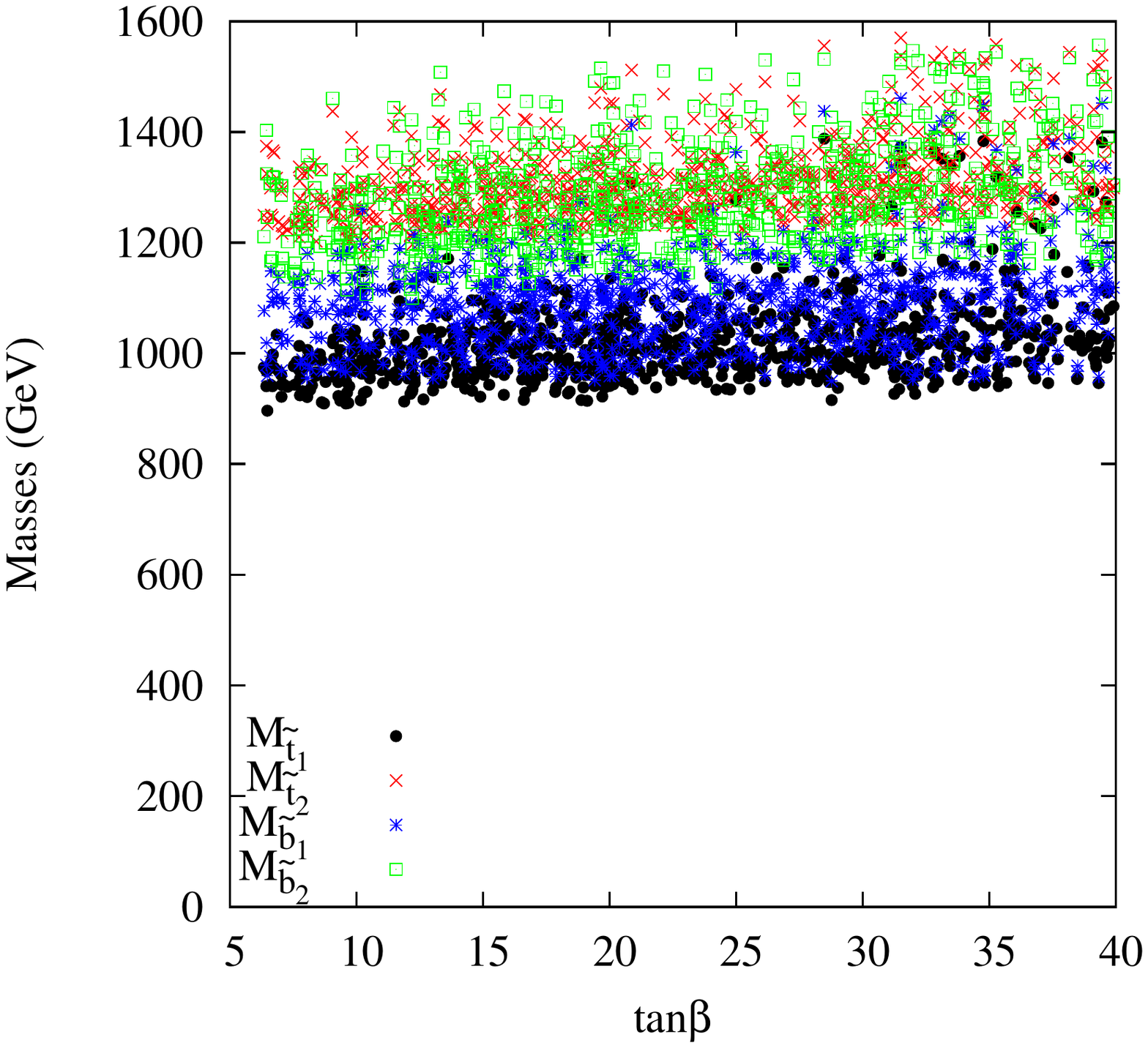}
\end{array}
$
\vskip-0.3in
\caption{The low lying mass spectrum of LRSUSY model as a function of 
$\tan\beta $  for 
$\mu ( gg \to H, H \to \gamma \gamma )= 1\pm 0.1 $ at 14 TeV LHC with
$300~{\rm fb}^{-1}$ luminosity. The left and right panels show weakly and
strongly interacting particle masses, respectively.}
\label{fig:futurerun}
\end{figure}

At the next run, LHC at 14 TeV with $300~ {\rm fb}^{-1}$ luminosity, the decay $H \to \gamma \gamma $ is expected to be measured with an accuracy of $10\%$ \cite{ATLAS004,CMS006}. With such precision,  our restrictions for the parameter space of the LRSUSY model will tighten considerably.   In anticipation, we illustrate in Fig. \ref{fig:futurerun} for BP3 the range of low lying LRSUSY weakly (left panel) 
and strongly (right panel) interacting masses as a function of $\tan\beta $ imposing the signal strengths $\mu (gg \to H \to \gamma \gamma) $ to be
measured with a precision of $10\%$. Among weakly interacting particles (charged Higgs bosons, charginos and staus), besides the LSP, the 
lowest lying particle is the  doubly charged scalar, whose mass 
can be as light as (200-300) GeV, while the heaviest one is the 
$\tilde \tau_2 $ with the mass in the range (450-600) GeV.  All these particles
will be easily accessible at the early run of the 14 TeV LHC. On the
other hand most of the strongly interacting particles (stop and sbottom) are heavier,  falling  
in the (1-1.4) TeV mass range. From these two figures, it clear that the mass
spectrum is almost independent of variations in $\tan\beta $. Note
that we did not use measurements on other Higgs signal channels as they have 
uncertainties larger than $20\%$. 

%%%%%%%%%%%%%%%%%%%%%%%%%%%%%%%%%%%%
\section{Conclusions}
\label{sec:conclusion}

In this paper we studied the implications of different Higgs signal
strengths as measured at the LHC on the parameter space of the 
left-right supersymmetric model (LRSUSY), especially on the Higgs, chargino and neutralino and scalar lepton sector. 
The LRSUSY models are based on by enlarging the standard model (SM) 
gauge group to $SU(3)_C\otimes SU(2)_L\otimes SU(2)_R\otimes U(1)_{B-L}$.
The gauge structure and particle contents of this model are such that it 
can generate the tiny masses for neutrinos as well as solve the strong 
and the EW CP problems. This model predicts a plethora of new particles,
among them the most important ones are singly and 
doubly charged Higgs bosons and higgsinos, which play crucial role in the
one-loop mediated Higgs boson decay in $\gamma \gamma $ and $Z\gamma $
channels. We presented a complete description of this model and chose 
certain benchmark points by fixing some basic parameters of the model,
while scanning over some other relevant parameters for the study of the
Higgs bosons decay patterns. It turns out that the lightest Higgs boson,
whose mass is close to 125 GeV and the lighter charged scalars are mostly components of the 
 Higgs $SU(2)$ bidoublet, while the right handed $SU(2)$ triplet yields the  doubly charged scalar. In our analysis we assumed that
the decay $H^{++} \to \tau^+\tau^+$ dominates, while other decay modes
are negligible,  and this allowed us to adopt the lower limit of 200 GeV for the
doubly charged Higgs boson masses, in agreement with limits obtained at the LHC. 

We estimated several Higgs signal strengths, defined as $R_{XX}$, 
in this model. For these,  we selected a particular benchmark point, namely
the BP3 and commented on changes expected by adopting BP1 or BP2 benchmarks. For the BP3 benchmark, we calculated explicitly the  
$R_{XX}$ values and then compared with the experimental values quoted by both
ATLAS and CMS for $\sqrt{s} = 7 $ TeV and 8 TeV at  $2\sigma $ precision.
We found that sufficient  LRSUSY model parameter space survives within the 
$2\sigma $ limit. To illustrate our findings, we considered either CMS or ATLAS results. Among the
different channels, we emphasized mainly $R_{WW^{*}}$ and $R_{\gamma \gamma }$
results, as these two are the most accurate and thus provide the most stringent limits on the parameter space.

In this scenario the one-loop mediated process, like
$H \to \gamma \gamma $ and  $H \to Z \gamma $, receive new 
contributions from doubly charged scalars and doubly charged higgsinos in 
addition to other supersymmetric particles. It is interesting to note that 
here the doubly charged scalar contribution is less than that of the
singly charged scalars in both $H \to \gamma \gamma $ and $H \to Z \gamma $
processes, simply because the $h-H^{\pm\pm}-H^{\mp\mp}$ coupling is weaker
than $h-H^{\pm}-H^{\mp}$ coupling, with a relative suppression of 
$\approx 1/20 $ which makes 
the doubly charged contribution substantially smaller than singly charged one.
This behavior is completely opposite to  Type-II Seesaw models, where 
the largest contribution to these one-loop processes come from the 
virtual exchange of doubly charged scalars.  Hence, perhaps this
feature can be used to distinguish this model from other models which also 
include doubly charged scalars. 

We also showed correlations (anti-correlations) among different $R_{XX}$ values
of the Higgs signal strengths by taking into account both the ATLAS and CMS
experimental results at $1\sigma $ and $2\sigma $ level. In particular, we found a nice
correlation between $R_{VV^\star}$ and $R_{\gamma \gamma }$, with $V=W^\pm, Z$. 
Our model predictions for these Higgs signal strengths in the  BP3 benchmark 
showed good agreement with both the ATLAS and CMS at $2\sigma $ level, each of which 
 could be matched individually, but not simultaneously, at $1\sigma $ level. This effect can be understood from the large difference between
the experimental central values and the corresponding error bars. 

The Higgs boson total width comes mainly from the Higgs partial 
decay width into $b \bar b$ final states, as expected, and this decay width is responsible
in controlling patterns of $R_{\gamma \gamma}$. We have showed that 
there is clear anti-correlation between $R_{\gamma \gamma}$ and $R_{bb}$ as
well as between $R_{\tau \tau} $ and $R_{bb}$. This is explained from the
fact that any increase in the partial width of $H \to b \bar b$ would lead
to a suppression in the partial widths of the Higgs boson into other channels, namely
into $\gamma \gamma $ and $\tau^+ \tau^- $ final states.  Once again for 
benchmark BP3, we showed that clear overlap regions are allowed at  
$2\sigma $ level from both the ATLAS and CMS experimental data on $R_{\gamma\gamma},
R_{bb} $ and $R_{\tau \tau}$. 

Assuming the fixed parameters from Table 1 for BP3, we finally predicted the low lying weakly and strongly interacting mass 
spectrum for the LRSUSY model if the decay width $H \to \gamma \gamma $ would be measured 
at the level of $10\%$ at 14 TeV LHC run with $300~{\rm fb}^{-1}$ luminosity, as it is expected.
We hope that some of these low lying particles will be seen and explored at the 
early run of the 14 TeV LHC. 
%%%%%%%%%%%%%%%%%%%%%%%%%%%%%%%%%%%%

\begin{acknowledgments}
D.K.G. and I.S. would like to acknowledge the hospitality provided by 
the High Energy Division, the Abdus Salam International Centre for Theoretical 
Physics (ICTP) and RECAPP, HRI where part of this work was done. The work of S.K.R. was partially supported by 
funding available from the Department of Atomic Energy, Government of India, for the Regional
Centre for Accelerator-based Particle Physics (RECAPP), Harish-Chandra Research Institute (HRI). 
D.K.G. and S.K.R. would also like to thank the University of Helsinki and the Helsinki Institute of Physics, where
this work was started and a major part of it was done and written. M.F. thanks the groups at HRI, IACS, India, and the University of Helsinki and the Helsinki Institute of Physics, and NSERC for partial 
financial support under grant number SAP105354.  
K.H. and H.W. acknowledge support from the Academy of Finland (Project No. 137960).
\end{acknowledgments} 

%\bibliographystyle{JHEP}

%%%%%%%%%%%%%%%%%%%%%%%%%%%%%%%%%%%%%%%%

\end{document}